\documentclass[12pt,preprint]{aastex}
\pdfoutput=1
\usepackage{emulateapj5}
\usepackage{natbib}
\usepackage{onecolfloat}
\usepackage{apjfonts}
\usepackage{subfigure}

\newcommand{\rfive}{$R_{500}$} 
\newcommand{\pthird}{$P_3/P_{0}$} 
\newcommand{\hinv}{$h^{-1}$} 
\newcommand{\minv}{$h^{-1} M_{\odot}$} 

\newcommand{\dist}{\emph {disturbed}} 
\newcommand{\reg}{\emph {regular}}
\newcommand{\chandra}{\emph {Chandra}} 

\newcommand{\bdm}{\begin{displaymath}} 
\newcommand{\edm}{\end{displaymath}}
\newcommand{\beq}{\begin{equation}} 
\newcommand{\eeq}{\end{equation}} 
\newcommand{\beqnarr}{\begin{eqnarray}}
\newcommand{\eeqnarr}{\end{eqnarray}}
\newcommand{\bit}{\begin{itemize}} 
\newcommand{\eit}{\end{itemize}} 
\newcommand{\ben}{\begin{enumerate}} 
\newcommand{\een}{\end{enumerate}}
\newcommand{\bfi}{\begin{figure}[htb]} 
\newcommand{\bpfi}{\begin{figure}[p]}
\newcommand{\barr}{\begin{array}}
\newcommand{\earr}{\end{array}}
\newcommand{\bec}{\begin{center}}
\newcommand{\eec}{\end{center}}
\newcommand{\bs}{\begin{sideways}}
\newcommand{\es}{\end{sideways}}

\shorttitle{X-ray morphological estimators for galaxy clusters}
\shortauthors{Rasia et al.}
\begin{document}
\twocolumn[%
\title{X-ray morphological estimators for galaxy clusters}
\author{
E.~Rasia\altaffilmark{1},
M.~Meneghetti\altaffilmark{2,3},
S.~Ettori\altaffilmark{2,3}
}
\altaffiltext{1}{Department of Physics, University of Michigan, 450 Church St., Ann Arbor, MI  48109, rasia@umich.edu}
\altaffiltext{2}{INAF, Osservatorio Astronomico di Bologna, v. Ranzani 1, I-40127, Bologna, Italy, massimo.meneghetti@oabo.inaf.it}
\altaffiltext{3}{INFN, Sezione di Bologna, viale Berti Pichat 6/2, I-40127, Bologna, Italy, stefano.ettori@oabo.inaf.it}

\begin{abstract}
The classification of  galaxy clusters according to their X-ray appearance is a powerful tool to discriminate between regular clusters (associated to relaxed objects)  and disturbed ones (linked to dynamically active systems). The compilation of  the two subsamples is a necessary step both for cosmological studies - oriented towards spherical and virialized systems- and for astrophysical investigations - focused on phenomena typically present in highly disturbed  galaxy clusters such as turbulence, particle re-acceleration, magneto-astrophysics .

In this paper, we review several morphological parameters: asymmetry and fluctuation of the X-ray surface brightness, hardness ratios, X-ray surface-brightness concentration, centroid shift, and third-order power ratio. We test them against 60 \chandra-like images obtained from hydrodynamical simulations through the X-ray Map Simulator and visually classified as regular and disturbed. 

The best performances are registered when the parameters are computed using the largest possible region (either within $R_{500}$ or 1000 kpc). The best indicators are the third-order power ratio, the asymmetry parameter, and the X-ray-surface-brightness concentration. All their combinations offer an efficient way to distinguish between the two morphological classes achieving values of purity extremely close to 1. A new parameter, $M$, is defined. It combines the strengths of the aforementioned indicators and, therefore, resulted to be the most effective parameter analyzed. 

\end{abstract}
]

\section{Introduction}\label{sec:intro} 


For over two decades, the X-ray-galaxy-cluster community has been searching for methods to suitably quantify the dynamical status of clusters based on their X-ray morphology \citep{jones&forman92}.
Initially, the morphological classification was mostly conducted by eye and strictly connected to the presence and characterization of substructures. For example, the morphological classes identified by \cite{jones&forman92} were: `single',`double', `primary with small secondary', `complex', `elliptical' (referred to the X-ray contours), `off-center' (either presenting a difference between the centers in optical and X-ray or showing an X-ray tail extended only in one sector off the X-ray peak), and `galaxy' (when the X-ray emission is dominated by the central galaxy emission). A very similar classification was proposed by \cite{baier.etal.96} who compared the substructures detected in X-ray with those known in optical wavelengths \citep[see also][]{kolokotronis.etal.01}. These visual estimates appeared to be robust \citep{jones&forman99} when compared with a more quantitative measurement of X-ray-morphology disturbance such as the power ratios \citep{buote&tsai95, buote&tsai96}, or axial ratios \citep{mohr.etal.93}.

Since its beginning the detection of X-ray substructures  has been  associated with dynamically active systems \citep[see also][]{slezak.etal.94,gomez.etal.97,rizza.etal.98,kolokotronis.etal.01,schuecker.etal.01}. Indeed, as demonstrated through hydrodynamical simulations,  a significant long time, spanning from 2Gyr to 4 Gyr, is required to re-establish a situation of equilibrium  after a major merger \citep{lacey&cole93,roettiger.etal.96,roettiger.etal.97,poole.etal.06,nelson.etal.12}. However, even if the presence of substructures is certainly indicative of a dynamically active system \citep{richstone.etal.92} the reverse is not always true. A cluster presenting either a strongly elliptical shape \citep{buote&tsai96,pinkney.etal.96,plionis02} or a variation of its X-ray centroid \citep{mohr.etal.95} is very likely an active cluster even if it has not recently interacted with a merging massive system and, therefore, does not present any substructure. Moreover, the visual classification  becomes challenging in prospect of the large number of data provided by future survey. A need for more robust indicators that can quantify objectively even small deviations from a perfectly regular and spherically-symmetric emission is, therefore, highly desirable. 

Among the various methods proposed, we recall some based on wavelets analysis \citep{pierre&starck,slezak.etal.94}, on the Minkowski functionals \citep{beisbard.etal.01}, on axial ratios \citep{mohr.etal.93}, on centroid-shift \citep[e.g.][]{mohr.etal.95}, and on power ratios \citep[e.g.][]{buote&tsai95}.
Several morphological indicators,  at the beginning, were used to test cosmological model  through a comparison of the X-ray morphological measurements performed in X-ray cluster surveys and in hydrodynamical simulated samples \citep{evrard.etal.93,jing.etal.95,dutta.etal.95,mohr.etal.95,tsai&buote,crone.etal.96,buote&xu97,thomas.etal.98,buote98,valdarnini.etal.99,suwa.etal.03,ho.etal.06}. Their usage in this context, has been significantly reduced recently as they are not as robust as expected for the level of precision required by current and up-coming cosmologically-driven surveys. However, a way to classify the X-ray appearance is still of great interest to generate subsamples of either very disturbed or very relaxed objects.

The first category provides the fundamental material for astrophysical investigations of many microphysical processes. Mergers are activating turbulence in the intracluster medium \citep{hallman.etal.11,vazza.etal.12}, in particular in the sloshing core \citep{zuhone.etal.12,roediger.etal.12}. The detection and statistics of cold fronts formed from the sloshing gas are a direct test of presence of thermal conduction \citep{zuhone.etal.12a}. Studies of cold-front edges will allow also to estimate the level of viscosity of the fluid through detection (or no-detection) of Kelvin-Helmotz instabilities \citep{markevitch&vikhlinin}.  Furthermore, merging clusters show correlation between the X-ray emission and the presence of giant radio halos \citep{buote01,cassano.etal.10, hallman.etal.11}. In this respect, the creation of highly-disturbed cluster sample will enable the examination of particle acceleration mechanisms resulting in non-thermal radio emission \citep{mann.etal.12}.

Regular clusters, instead, can be used to calibrate scaling relations. Since their X-ray mass derived by assuming hydrostatic equilibrium is believed to be more robust \citep{rasia.etal.04,rasia.etal.06,nagai.etal.07,piffaretti&valdarnini,lau.etal.09} and closer to the gravitational lensing masses \citep{meneghetti.etal.10,rasia.etal.12, chon.etal.12,mahdavi.etal.12}.
Furthermore, regular systems have a reduced scatter in scaling relations \citep[e.g.][]{rowley.etal.04,croston.etal.08,yang.etal.09,rasia.etal.11}. 
Irregular systems, on the contrary, can be used in scaling relation only if their level of substructure is known  (see recent method proposed by \citealt{andrade.etal.12}). Indeed, even if typically irregular systems at fixed X-ray temperature tend to be less luminous   \citep{torri.etal.04,kay.etal.07,hashimoto.etal.07,chon.etal.12, andrade.etal.12} and less massive  \citep{ohara.etal.06,valdarnini.etal.06,ventimiglia.etal.08,ventimiglia.etal.12,chon.etal.12, andrade.etal.12}, the parametrization of the degree of substructures can be used to correct their positions in the $L-T$ or $M-T$ planes \citep{ventimiglia.etal.08}.

Morphological parameters will be therefore useful to categorize the large amount of data that will be collected from future surveys and to parametrize their X-ray morphology. With this perspective in mind, for the first time,  we test several morphological indicators against the same ensemble of synthetic X-ray data. The final goal is to identify the parameter(s) that most efficiently allow for the classification of regular and irregular objects. We establish these classes by visual analysis of the images, i.e. following a pure observational approach. At first, we distinguish three classes: \reg, \dist, and `mix'. In the last Section we subdivided the last category into {\it semi-regular} and {\it semi-disturbed} systems.

The simulated sample and the X-ray synthetic catalogue are presented in Section 2, while the morphological parameters are introduced in Section 3. Each parameter is tested against the visual classes of \reg\, and \dist\, objects (Section 4). To allow for statistical evaluations, we extend the analysis to the whole sample when studying the improvement generated by combining two parameters (Section 5).
Conclusions follow in Sections 6. In the appendix, we present the soft  band images of all our clusters.

%
%
%
%
%
%
%
%
%
%
%

\section{Simulations}\label{sec:sim} 

The 60 X-ray images analyzed in this paper reproduce \chandra-like event files of 20 simulated clusters observed along three lines-of-sight each. This set of synthetic observations have been also used in Rasia et al. (2012, hereafter R12) and Rasia et al. (in prep.). Here we briefly describe the main characteristics of the simulations and of the X-ray catalogue. Further details can be found in the aforementioned publications.

\subsection{Simulated clusters}
The simulated clusters used in this work are obtained as follows. Twentynine Lagrangian regions have been selected from a large (1 Gpc$^3 h^{-3}$) DM-only cosmological box. Twenty five of these regions are centered on the most massive objects of the parent simulation while the remaining are around group-size halos. The regions have been subsequently re-simulated with an increased resolution in mass following the Zoomed Initial Condition technique \citep{tormen.etal.97,bonafede.etal.11} and considering various recipes for the intra-cluster medium physics  \citep{fabjan.etal.11,killedar.etal.12}. In this work, as in R12, we consider only the  CSF (cooling-star formation-feedback) runs. The finest mass resolutions for the dark matter and gas particles are, respectively, $m_{dm}= 8.47 \times 10^8$ \minv and $m_{gas} = 1.53 \times 10^8$ \minv. The Plummer equivalent force used to compute the gravitational interaction in the high resolution regions is kept fixed to $\epsilon=5$ $h^{-1}$ kpc in comoving units from redshift $z=60$ to $z=2$. Thereafter, the same value is kept constant in physical units. The minimum SPH smoothing length  allowed is  $0.5 \times \epsilon$.

We assume a flat $\Lambda$CDM model with cosmological parameters in agreement with WMAP7 constraints \citep{komatsu.etal.11}: 
$\Omega_M = 0.24$ for the matter density parameter, $\Omega_{\Lambda}=0.76$ for the dark energy, $\Omega_{bar} = 0.04$ for the baryon density, $H_0 = 72$ km s$^{-1}$ Mpc$^{-1}$ for the Hubble constant at redshift zero, $n_s = 0.96$ for the primordial spectral index, and $\sigma_8 = 0.8$ for the normalization of the power spectrum on a scale of 8 $h^{-1}$ Mpc.

The hydro-resimulation are performed using the TreePM - SPH GADGET-3 code an algorithmically improved version of the original code GADGET-2 \citep{springel.etal.05}.  The physics implemented in the resimulations analyzed in this paper includes a model for heating and cooling from an evolving but spatially-uniform UV background \citep{wiersma.etal.09}. The star formation is modeled using the recipe of \cite{springel&hernquist}: when the gas particle reaches a density above a certain threshold it is treated as a multi-phase particle with the  hot and cold phases in pressure equilibrium. The last one serves as reservoir of star formation. The intracluster medium is enriched by various metals expelled by different stellar populations \citep{tornatore.etal.07}. The galactic winds mimicking supernovae explosions have a  velocity of 500 km s$^{-1}$.

The simulation outputs considered here refer to redshift  $z=0.25$. In this way, all objects have their $R_{500}$\footnote{Radius of the sphere whose density is 500 times the critical density at the cluster redshift.} within the field of view. The total mass enclosed in $R_{200}$ ($R_{500}$) spans from 4 to 15 (from 2 to 11) times $10^{14}$ \minv .
 Other works in literature \citep{poole.etal.06,jeltema.etal.05,jeltema.etal.08,maughan.etal.08} have shown how the hierarchical formation of structure implies a logical change of the distribution of the morphological parameters: a set of high-redshift clusters will be an average more disturbed. We remind that, in this paper, we aim at testing the efficiency of the investigated parameters rather than furnishing precise statistical indications for the distributions of the parameters over time. Selecting a medium redshift as 0.25 is a suitable choice for  our investigation.

\subsection{X-ray Catalogue}

To facilitate the post-processing  on the X-ray images we excluded {\it a priori} some particles before creating the event files. Namely, we omitted the coldest and densest particles satisfying the following condition: $T_p < 3 \times 10^6 \rho^{0.25}_{\rm gas,p}$. These particles exist in the simulation as consequence of the overcooling phenomenon that typically affects hydrodynamical simulations with cooling and star formation \citep[see discussion in][]{borgani&kravtsov}. The normalization value of this relation was set empirically by looking at the density and temperature of all particles of each of the twenty clusters. The relation was defined such to separate the over-cooled particles from the thermalized-cluster particles. The exponent of the density, 0.25, is linked to the politropic index of the simulations (see Appendix A of R12 for details).

The event files are created thanks to the X-ray MAp Simulator (X-MAS) code \citep[presented in][]{gardini.etal.04,rasia.etal.08} to reproduce \chandra-like observations. We  consider the ancillary response function (ARF) and redistribution matrix function (RMF) of the aim-point of ACIS-S3 throughout the field of view of size equal to 16 arcmin (equivalent to 2561 $h^{-1}$ kpc accordingly to the assumed cosmology and  considered redshift, $z=0.25$). For the galactic absorption, we adopt a WABS model fixing the column density at $N_H=5 \times 10^{-20}$ cm$^{-2}$. The X-ray emission, instead, has been reproduced using a MEKAL model in the XSPEC package \citep{arnaud96} and assuming a constant metallicity equal to 0.3 times the solar metallicity measured by \cite{anders&grevesse}. 
The exposure time was set to 100 ksec.

The imaging analysis conducted in this paper, uses packages developed within CIAO  \citep{fruscione.etal.06} and IDL subroutines. Contrarily to what done in R12, in this work, we do not exclude any region or substructure or merging clump within \rfive. The goal of the previous work  was, indeed, to measure the X-ray hydrostatic mass and therefore we were interested to characterize the regularity of the portion of the gas used to compute the total mass.  This paper, instead, aims at characterizing the overall X-ray morphology. 

\section{Morphological Estimators}\label{sec:morpho} 

In this Section we introduce the morphological parameters investigated presenting the original definition as well as their modified version adopted  in this paper. The first three parameters (Asymmetry, Fluctuation and X-ray Concentration) are strongly connected to the definition of the `CAS' parameters provided by \cite{conselice03}. The original work by Conselice was dedicated to optical images. Specifically, the Author was looking for a classification method  allowing the connection between the structural appearance of the stellar light and the galactic formation history.
The acronym `CAS' stands for Concentration of light, Asymmetry, and Spatial frequency clumpiness parameter. Very similar parameters have been introduced in relation to X-ray images throughout the years with modified nomenclature: X-ray surface brightness concentration, asymmetry, and fluctuation, respectively \citep[e.g.][]{cassano.etal.10, okabe.etal.10, zhang.etal.10}.

Furthermore, in addition to the hardness ratio parameter, we test other three indicators (the centroid shift, the third-order power ratio,  and its maximum). These have been extensively used in X-ray works on clusters of galaxies \citep[e.g.][R12]{mohr.etal.95,buote&tsai95,buote&xu97,jeltema.etal.05, bauer.etal.05,maughan.etal.08,ventimiglia.etal.08,jeltema.etal.08, andersson.etal.09,hallman.etal.10,cassano.etal.10,hallman.etal.11, boehringer.etal.10,weissmann.etal.13}. Specifically, two recent works, have also addressed the issues of Poisson noise and X-ray background in evaluating these parameters and offered strategies to adopt in real observations \citep{boehringer.etal.10,weissmann.etal.13}.

\paragraph{Fluctuation parameter, $F$.}
 The fluctuation parameter emphasizes the presence of peaks (valleys) of high (low) X-ray flux over the image. Various similar definitions have been historically used. We prefer an expression similar to the asymmetry parameter (see below): $F$ is obtained by subtracting a smoothed image, $B$, from the original one, normalizing by the original image: 
\beq
  F=\frac{\Sigma(|I-B|)}{\Sigma(I)}.
\label{eq:f}
\eeq 
\cite{conselice03} used a similar parameter, called $S$, for High-Spatial frequency clumpiness that, however, considered only the peaks of the surface brightness. Indeed, in the computation of the sum in  Eq.~\ref{eq:f} the absolute value was left out and all negative differences of ($I-B$) were discarded. The parameter was computed without the absolute value also by the Locuss collaboration \citep{okabe.etal.10,zhang.etal.10} but the negative deviations were kept in the sum with their sign, in a way that positive enhancements were eliminated by negative depressions. Their smoothed map was obtained by assuming a Gaussian filter of FWHM equal to 2$^{\prime}$ (equivalent to 400 kpc at z=0.2). Accordingly to this assumptions the fluctuation parameters varied from 0 to 0.14 and 0.05 was considered the boundary dividing \reg\, and \dist\, objects.

We derived three smoothed maps by varying the Gaussian-filter FWHM from $2^{\prime}$ to $16^{\prime \prime}$, and to 8$^{\prime \prime}$ to test the effect of the smoothing on the results. The associated fluctuation parameters are, respectively, called $F_1$, $F_2$, and $F_3$.

\paragraph{Asymmetry parameter, $A$.}
One of the most powerful indicator of the morphology of a galaxy is the asymmetry parameter \citep[e.g.][]{schade.etal.95} obtained  by summing the residuals between an image, $I$, and its 180$^{\circ}$ rotated counterpart, $R$, normalized by the original image:
\beq
A=\frac{\Sigma (|I-R|)}{\Sigma(I)} 
\label{eq:a}
\eeq 
In this fashion, all asymmetries non perfectly aligned with the diagonals are emphasized.

The Locuss collaboration applied this definition to soft X-ray images obtained by XMM-Newton \citep{zhang.etal.10,okabe.etal.10}. The images were extracted in the soft band ([0.7-2.] keV), corrected for the flat field, point-source-subtracted, and binned by $4^{\prime \prime} \times 4^{\prime \prime}$. These works pointed out that the clusters showing an high level of asymmetry are presenting as well a discrepancy between the X-ray centroid and weak-lensing center. This last feature clearly denotes a recent dynamical activity. The Authors identified as \dist\, clusters all objects having $A>1.1$. The full range of the asymmetry parameter for their sample spanned from 0.07 to 1.5.

In this work we consider three different asymmetry parameters. Namely, the image $R$ is rotated by $180^{\circ}$, $A_{\rm rot}$, $ii)$ flipped along the $x-$axis, $A_x$, and $iii)$ flipped along the $y-$ axis, $A_y$, with both axis passing through the X-ray centroid. Finally, per each cluster, we also consider the maximum among the three asymmetry parameters ($A_{\rm max}$). With this procedure, we avoid any possible underestimate of non-asymmetry present along a favorite direction.

\paragraph{Light Concentration, $c$.}

Analogously to the definition of `light concentration' by  \cite{conselice03} \citep[see also][]{bershady.etal.00} a specific parameter measuring the concentration of the X-ray-emission has been defined by \cite{cassano.etal.10}. This is as:
\beq
c_{\rm [kpc]}=\frac{S(r<100 kpc)}{S(<500 kpc)}.
\label{eq:csb}
\eeq
A similar definition  was used by \cite{santos.etal.08} as an indicator for the presence of cooling core systems at high redshift: $c=S(r< 40 {\rm kpc})/S(r<400 {\rm kpc})$.  The radii of 40 and 400 kpc were chosen to maximize the separation of concentration values between cool-core (cc) and non cool-core (ncc) clusters \citep[see also][]{semler.etal.12}. However, simulations with radiative cooling and not powerful feedback such as that provided by active galactic nuclei are still unable to reproduce the statistic of cc versus ncc systems \citep[see the review by ][ regarding the over-cooling problem affecting simulations]{borgani&kravtsov}. For this reason, we select a larger region to represent the core. \cite{hallman.etal.11} noted the limitations of considering apertures at fixed physical radii  in the study of large samples of clusters at different redshifts. They proposed the usage of apertures with radii expressed in units of $R_{500}$ as done in \cite{kay.etal.07}. Taking into account this comment, in addition to the definition presented in Eq.~\ref{eq:csb} we test another definition of the concentration:
\beq
 c_{\rm [R_{500}]} = \frac{S(r< 0.2\times R_{500})}{S(r<R_{500})}. 
\eeq

\cite{cassano.etal.10} studying 32 clusters observed with \chandra\, (with redshift between 0.2 and 0.4 and X-ray luminosity above $10^{44}$ erg s$^{-1}$) proposed $c_{\rm kpc}=0.2$ (median value of their distribution spanning from $\sim$ 0.05 to $\sim$ 0.7) as the dividing line from \reg\, and \dist\, systems.
\paragraph{Hardness ratio indicators, $H$.}
 In the hierarchical structure formation scenario larger system form through accretion of smaller structures. In this picture, a merger only sporadically involves two objects with the same mass and, most likely, happens between two or more systems presenting a large spread in mass that can be translated into a spread in temperature. A good variety of movies based on simulations clearly shows how the thermalization process of the intra-cluster gas is continuously challenged by the incursion of smaller and colder blobs \footnote{See, for example, the movies at the following link http://www.mpa-garching.mpg.de/galform/data\_vis created by Klaus Dolag}.  The multi-temperature phase of the ICM of a merging object can be enhanced by taking the ratio between images obtained in the soft band (such as [0.3-1.5] keV where the cold gas is strongly emitting) and  in the hard band (such as [1.5-7.5] keV where the hot gas has more power). This technique, recently used by \cite{gitti.etal.11} to identify cold filaments in {\it Hydra A} cluster, has a long history of usage to identify temperature structures \citep[e.g.][]{allen&fabian,ettori.etal.98}.

In R12, we considered two hardness ratio 
\beq
H_1 = \frac{\Sigma (H-S)}{\Sigma(S)} \, \, \, \, \, {\rm and} \, \, \, \, \, H_2 = \Sigma (H/S) 
\label{eq:h}
\eeq
where $H$ and $S$ are the images in the aforementioned hard and soft energy bands, respectively. 
In this paper, we focus only on the first definition that has been proved to be more effective and  we also consider another ratio defined with $S$ and $H$ extracted in the soft energy band [0.7-2] keV and in the hard energy band [2-7] keV, respectively. Before computing the ratios, we applied to both images a Gaussian filter of FWHM equivalent to $2^{\prime}$, 16$^{\prime\prime}$, and 8$^{\prime\prime}$.


\paragraph{Centroid shift, $w$.} 
This measure, as the name suggests, assesses how much the centroid of the X-ray surface brightness moves when the aperture used to compute it decrease from a certain $R_{max}$ to smaller radii. The presence of X-ray bright clumps can produce significant changes on the X-ray centroid unless they are distributed in a perfectly symmetric geometry with respect to the center. 
The parameter is defined as follow:
\beq
w=\frac{1}{R_{\max}} \times \sqrt{\frac{\Sigma(\Delta_i-<\Delta>)^2}{N-1}}
\label{eq:w}
\eeq
where $N$ is the total number of apertures considered and $\Delta_i$ is the separation of the centroids computed within $R_{\rm max}$ and within the $i^{th}$ aperture.
This parameter has been extensively used in literature \citep[e.g.][]{ohara.etal.06,poole.etal.06,maughan.etal.08,ventimiglia.etal.08,jeltema.etal.08,boehringer.etal.10,weissmann.etal.13} since it is an useful characterization of the dynamical state of the cluster \citep{mohr.etal.93}. The robustness holds also if the parameter is slightly differently defined \citep{thomas.etal.98, kay.etal.07}.
 We compute Eq.~\ref{eq:w} by varying the apertures from $0.15 \times$ $R_{\max}$, to $R_{\max}$ with a step of $ 0.05 \times R_{\max}$ and 
$R_{\max}$ varying from 30\% to 1 $R_{500}$.

The most recently chosen boundary between \reg\, and \dist\, objects present in literature is around 0.01. Specifically, $w=0.012$ for \cite{cassano.etal.10} who studied 45 luminous clusters observed with \chandra, $w=0.01$ for \cite{weissmann.etal.13} for 80 clusters observed with XMM-Newton, $0.02$ for \cite{ohara.etal.06} who analyzed 45 clusters from the ROSAT-PCPS images (thus with less spatial resolution than \chandra). The ranges reported in literature span from $\sim$ 0.001 to $\sim$ 0.15 \citep{poole.etal.06,ventimiglia.etal.08, jeltema.etal.08,cassano.etal.10,weissmann.etal.13} with some small departures on the limits depending on the $R_{max}$ chosen.

\paragraph{Power ratios, \pthird.} 
The power ratios, introduced by \cite{buote&tsai95}, mimic a multiple decomposition of the two-dimensional projected mass distribution inside a certain aperture, $R_{\rm ap}$. Instead of the mass, however, they are applied to the X-ray surface brightness images, $S$.

The generic $m$-order power ratio ($m>0$) is  defined as $P_m/P_0$ with
\beq
P_m=\frac{1}{2m^2R^{2m}_{\rm ap}}(a_m^2 +b^2_m) \ \ \ \ {\rm and} \ \ \ \  P_0=a_0 \ln(R_{\rm ap})
\eeq
where $a_0$ is the total intensity within the aperture radius. The generic moments $a_m$ and $b_m$ are expressed in polar coordinates, $R$ and $\phi$, and given by
\beq
a_m(r)=\int_{R^{\prime} \leq R_{\rm ap}} S(x^{\prime}) R^{\prime} cos(m \phi^{\prime}) d^2 x^{\prime},
\eeq
and
\beq
b_m(r)=\int_{R^{\prime} \leq R_{\rm ap}} S(x^{\prime}) R^{\prime} sin(m \phi^{\prime}) d^2 x^{\prime}.
\eeq

The quadrupole power $P_2$ refers to the ellipticity of the clusters, $P_3$ informs about bimodal distribution, $P_4$ is similar to $P_2$ but more sensible to smaller scales (indeed, $P_2$ is strongly correlated with $P_4$). The third order power ratio is the most suitable to identify asymmetries or presence of substructures. This is the parameter investigated in this paper. In literature, various aperture radii have been used both at a precise physical scale (either 0.5 Mpc or 1 Mpc) or at $R_{500}$. In this paper, we will test all these possibilities  as well as the new parameter described in the next paragraph. In most recent literature, the decimal logarithm of \pthird ranges from $\sim$ -8.7 to $\sim$ -5.7 \citep{jeltema.etal.05, jeltema.etal.08,cassano.etal.10, weissmann.etal.13} again depending on the maximum radius used for the calculation of the power ratios. The median values are typically log$_{10}$(\pthird)$=-7.2$ and the dividing limit between \reg\, and \dist\, system is set around  $-6.7$ and $-7$. \cite{weissmann.etal.13} proposed other two boundaries to select exclusively \reg\, and \dist\, objects: \pthird$<10^{-8}$ and \pthird$>5 \times 10^{-7}$, respectively.


\paragraph{Maximum of the third power ratio, \pthird$_{\rm max}$.} 

In a recent paper, \cite{weissmann.etal.13} discussed the limits of the third order power ratio computed within a fixed radius (either in physical units or in units of  $R_{500}$). Their claim is based on the consideration that the \pthird ~ of the Bullet cluster evaluated at $R_{500}$ has a value typical of a regular system ($\sim 10^{-7}$). This astonishing result is explained by the fact that the influence of the `bullet', located in a more central zone, is smeared out when considering the entire cluster volume. The evaluation of the third order power ratio at a fixed radius limits the power to discriminate between `regular' and `disturbed' objects. 
To fixed this problem,  \cite{weissmann.etal.13} proposed to consider the maximum of the third order power ratios estimated at various radius from 0.3 to 1 $R_{500}$ with step equal to $0.1 \times R_{500}$. We, therefore, compute all eight power ratios, and per each cluster consider
the maximum among them.

\section{Results}\label{sec:anal} 

Unless otherwise specified, all morphological parameters are measured  on the soft X-ray image binned $2^{\prime \prime} \times 2^{\prime \prime} $ at various ratios of $R_{500}$: 0.3, 0.5, 0.8, and 1. The efficacy is proved against the visual classification established after a careful process. 

\paragraph{Visual classification.}
In order to closely reproduce the observational approach we divide the sample into three classes: \reg, \dist, and `mix'. To avoid influence of the observer-bias, three different people looked at and rated the images. These were shown three times each and appeared in random order. For every image, we average the nine grades. These span from 0 (given to a spherical system without substructures) to 3 (very disturbed X-ray emission or clear evidence of large merging substructures). Strongly elliptical clusters or spherical clusters with minor substructures still in the outskirts (that have not crossed $R{500}$) and not interfering with the X-ray emission of the main system resulted typically rated from 1 to 2. All images with an average grade below or equal 1.5 are considered of \reg\, systems, while those with a grade above 2.5 are proper of \dist\, objects. The two subsamples are shown in Fig.~\ref{fig:reg} and Fig.~\ref{fig:dist} of the Appendix. All other objects fall into a class we call `mix' (Fig.~\ref{fig:mix}). As evident from the images, our \reg\, clusters are part of the `single', `elliptical' (even if not {\it strongly} elliptical), and `galaxy' categories proposed by \cite{jones&forman92} while the \dist\, system could be represented by their `complex', `double', `off-center' classes.
In next Section, we will study the efficiency and improvement of combining more estimators together and will calculate some statistical parameters as {\it purity} and {\it completeness}. For this reason, we will need to consider the whole sample. Consequently, we proceed to subdivide the `mix' class into {\it semi-regular} and {\it semi-disturbed}. The  limit of the averaged grade chosen to divide the two extended classes is set equal to 2.

\paragraph{Asymmetry and fluctuation parameters}

We investigate the efficiency of the asymmetry and fluctuation parameters in distinguishing among \reg\, and \dist\, clusters both individually and combining them together \citep{okabe.etal.10,zhang.etal.10}.

\begin{figure}[h]
\centering
\includegraphics[width=0.45\textwidth]{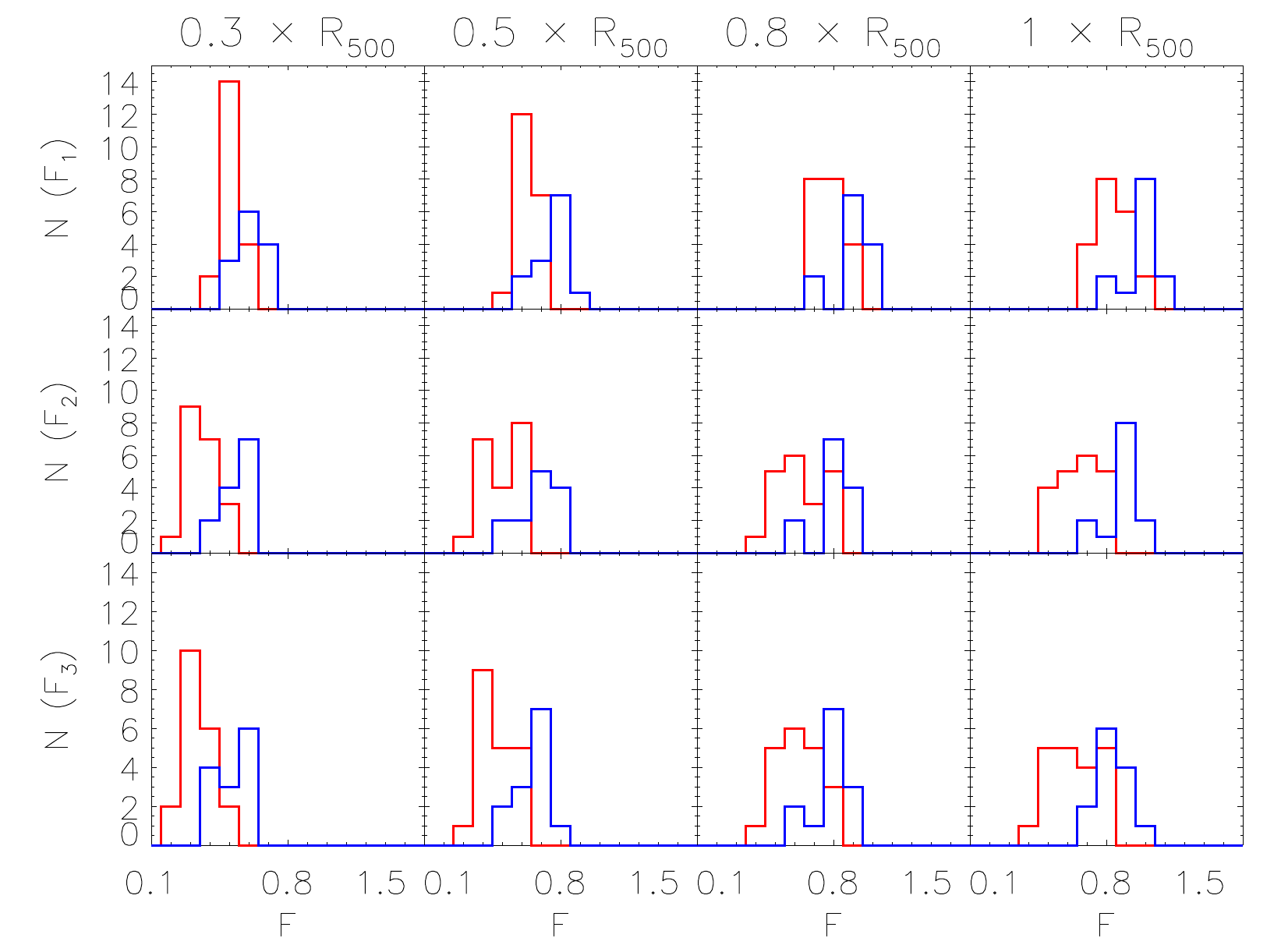}
\includegraphics[width=0.45\textwidth]{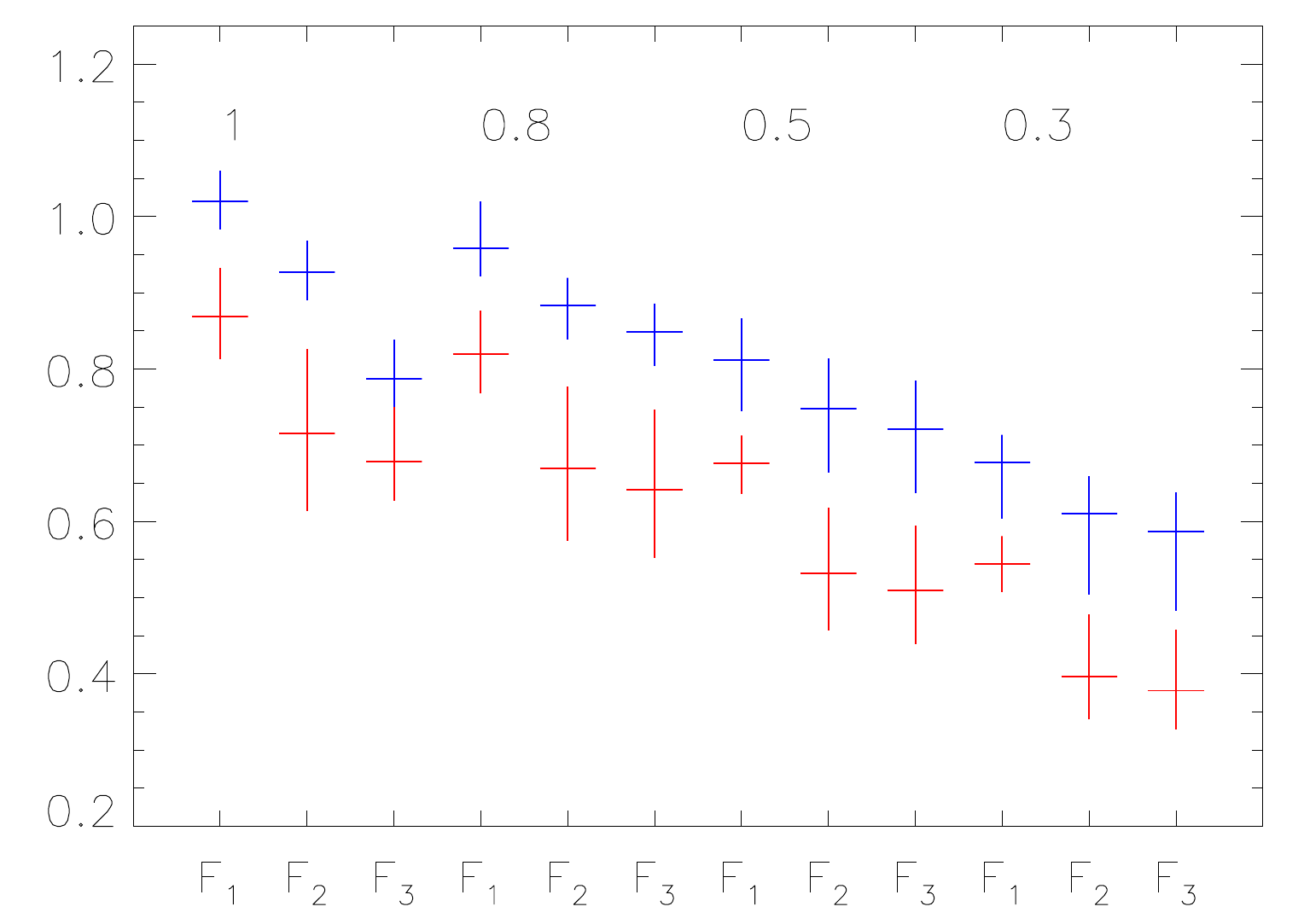}
\caption{Top panel: distribution of the {\it regular} (red) and {\it disturbed} (blue) clusters as a function of the fluctuation parameter, $F$. The smoothed image used in Eq.~\ref{eq:f} is derived with a Gaussian smoothing of FWHM equal to $2^{\prime}$ ($F_1$, upper panels), $16^{\prime\prime}$ pixel ($F_2$, middle panels), and $8^{\prime\prime}$ pixel ($F_3$, bottom panels).  The sum is computed over a circular region of radius specified at the top of the columns. Bottom panel: variation of the fluctuation parameter statistic (vertical values show the range between $q_1$ and $q_2$ of Table~1). The numbers listed within the Figure show the fraction of $R_{500}$ used in the sum of Eq.~\ref{eq:f}}. 
\label{fig:fluc}
\end{figure}


Per each of the three smoothed maps, we evaluate $F$ at four different radii, and then compare all results with the visual classification. In the left panel of Fig.~\ref{fig:fluc}, we show the histograms of $F_1, F_2, F_3$ for \reg\, (in red) and \dist\, (in blue) clusters (for clarity, we omit the `mix' class). The distributions of the whole sample, \reg\, sub-sample and \dist\, one are statistically characterized by the median, the first and third quartiles, reported in Table~\ref{tab:stat_f} and plotted in the right panel of Fig.~\ref{fig:fluc} (for the following parameters we will omit the figure since the information is already contained in the tables). We, further, apply a Kuiper-two-sample-test statistics to determine the probability that the two distributions are statistically different. For the numbers in this work (20 \reg\, clusters and 13 \dist\, ones) a probability $p$ below 0.05 implies a net distinction between the two distributions. 

We conclude that the fluctuation parameter is a good indicator of the X-ray morphology, in particular, when the measures are done at $R_{500}$ and the largest Gaussian filter is applied  ($F_1$ performs better than $F_3$ in all cases apart in the real central zone). Specifically, the limit of $F_1 ({\rm at ~} R_{500})=0.9$ effectively divided the two morphological samples. Measuring the parameter at the center of the cluster ($0.3 \times R_{500}$) is a good strategy to select \reg\, objects: when $F_1 < 0.6$ ($F_3<0.5$)  84\% (82\%) of the objects selected are \reg\, and the medians of the distributions are significantly apart.

\begin{table*}
\caption{Fluctuation and asymmetry parameter distributions (median,first and third quartiles) computed at various fraction of $R_{500}$ for the regular sample, all the clusters, and the disturbed sample. Last column refers to the Kuiper-two-sample-test probability.} 
\centering
\begin{tabular}{|c|ccc|ccc|ccc||c|}
\hline
$par$ &\multicolumn{3}{c|}{\reg} &\multicolumn{3}{c|}{all} &\multicolumn{3}{c||}{\dist} &  \\
 & $q^{\rm 1st}$ & med & $q^{\rm 3rd}$ & $q^{\rm 1st}$ & med & $q^{\rm 3rd}$  & $q^{\rm 1st}$ & med & $q^{\rm 3rd}$ &  $p$\\
\hline
$F_1 \ (R_{500})$                         &  0.81   &    0.87  &   0.93  &      0.80  &     0.90    &      1.01    &     1.01   &     1.02       & 1.06  & 0.008\\
$F_2 \ (R_{500})$                         &  0.61    &   0.72   &   0.83  &      0.65  &     0.76   &     0.91   &      0.90 &      0.93   &    0.97      &0.001 \\
$F_3 \ (R_{500})$                         &   0.33    &   0.38   &   0.46   &     0.63  &     0.73  &     0.87   &      0.48  &     0.59   &    0.64     & 0.001\\
& & & & & & & & & & \\
$F_1 \ (0.8 \times R_{500})$       &    0.77   &    0.82   &   0.88   &     0.76  &     0.84   &    0.95   &      0.94  &     0.96   &     1.02  &  0.012\\
$F_2\  (0.8 \times R_{500})$       &    0.58    &   0.67    &   0.78   &     0.60  &     0.72   &    0.85   &      0.84   &    0.88    &   0.92   & 0.012\\
$F_3 \ (0.8 \times R_{500})$       &    0.55    &   0.64    &   0.75   &     0.58  &     0.69   &    0.82   &      0.80   &    0.85   &    0.88    & 0.012\\
& & & & & & & & & & \\
 $F_1 \ (0.5 \times R_{500})$      &  0.64   &    0.68   &   0.71   &     0.64  &     0.68   &    0.76   &      0.74   &    0.81   &    0.87   &  0.012 \\
 $F_2 \ (0.5 \times R_{500})$       & 0.46    &   0.53    &   0.62   &     0.48  &     0.60   &    0.70   &      0.66   &    0.75   &    0.81   & 0.033\\
 $F_3\  (0.5 \times R_{500})$       & 0.44    &   0.51    &   0.59   &     0.46  &     0.57   &    0.67   &      0.64   &    0.72   &    0.78   & 0.033\\
& & & & & & & & & & \\
 $F_1 \ (0.3 \times R_{500})$       & 0.52   &    0.54    &  0.57    &    0.52  &     0.56   &    0.62  &       0.60   &    0.68  &     0.71    & 0.033\\
  $F_2\  (0.3 \times R_{500})$       & 0.34    &   0.40     &  0.48    &    0.36  &     0.46   &    0.54  &       0.50   &    0.61   &    0.66  & 0.004\\
  $F_3 \ (0.3 \times R_{500})$       & 0.33   &    0.38     &  0.46    &    0.34  &     0.45   &    0.52  &       0.48    &   0.59   &    0.64  & 0.001\\
& & & & & & & & & & \\
\hline
\hline
$A_r \ (R_{500})$                      &    0.84 &      0.93   &     1.08  &           0.87   &     1.06  &       1.17      &        1.29 &       1.30   &     1.34  &    2 E-4 \\
$A_x \ (R_{500})$                     &    0.83  &     0.93   &     1.08   &          0.87   &     1.07   &     1.18       &       1.19   &     1.30    &    1.37    &    6 E-4 \\
& & & & & & & & & & \\
$A_r \ (0.8 \times R_{500})$      &    0.80  &     0.89  &      1.06    &         0.83   &     1.02   &     1.10       &       1.23     &   1.25    &    1.29    &  2 E-4 \\
$A_x \ (0.8 \times R_{500})$      &   0.80   &    0.89  &      1.05     &        0.83   &     1.03    &    1.14       &       1.14      &  1.27    &    1.31    &  0.004 \\
& & & & & & & & & & \\
$A_r \ (0.5 \times R_{500})$      &    0.69   &    0.77   &    0.92    &         0.74  &     0.89    &   0.96   &           1.05   &     1.08   &     1.12   &   0.004 \\
$A_x \ (0.5 \times R_{500})$      &   0.68    &   0.77   &    0.90     &        0.73  &    0.88    &   0.99    &         0.99    &    1.10    &    1.12    &  0.020 \\
& & & & & & & & & & \\
$A_r  \ (0.3 \times R_{500})$      &   0.57   &    0.64  &     0.76      &       0.59  &     0.69    &   0.83   &          0.85  &    0.88  &     0.91   &   0.008 \\
$A_x \ (0.3 \times R_{500})$      &   0.55    &   0.64  &     0.75       &      0.59  &     0.72    &   0.84   &          0.83   &   0.88  &     0.92    &  0.021 \\
& & & & & & & & & & \\
$A_{\rm max} \ (R_{500})$                      &   0.85   &    0.94  &     1.10  &   0.89  &      1.09  &      1.18  &  1.29  &      1.34  &      1.37 &    8 E-5 \\
$A_{\rm max} \ (0.8 \times R_{500})$     &   0.81    &   0.90   &    1.06  &   0.85  &      1.05  &      1.14  &  1.25  &      1.28  &      1.31 & 3 E-4 \\
$A_{\rm max} \ (0.5 \times R_{500})$     &   0.70    &   0.78   &    0.93  &   0.75  &      0.91  &      0.99  &  1.06  &      1.11  &      1.14 &   0.004 \\
$A_{\rm_max}  \ (0.3 \times R_{500})$   &   0.57    &   0.66   &    0.78  &   0.60  &      0.74  &      0.85  &  0.85  &      0.91  &      0.95 &   0.008 \\
& & & & & & & & & & \\
\hline
\end{tabular}
\label{tab:stat_f}
\end{table*}

\begin{figure}[h!]
\centering
\includegraphics[width=0.45\textwidth]{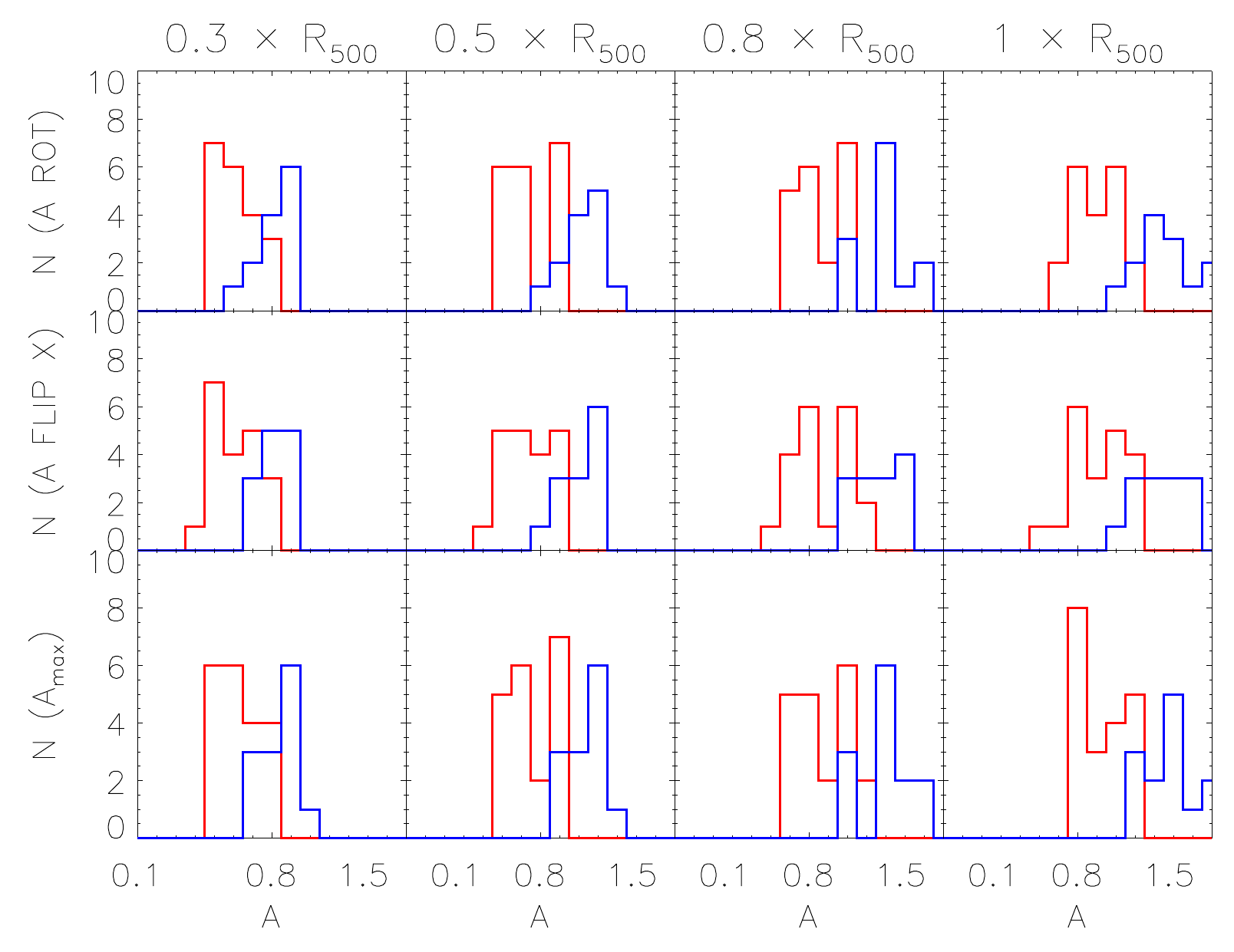}
\includegraphics[width=0.45\textwidth]{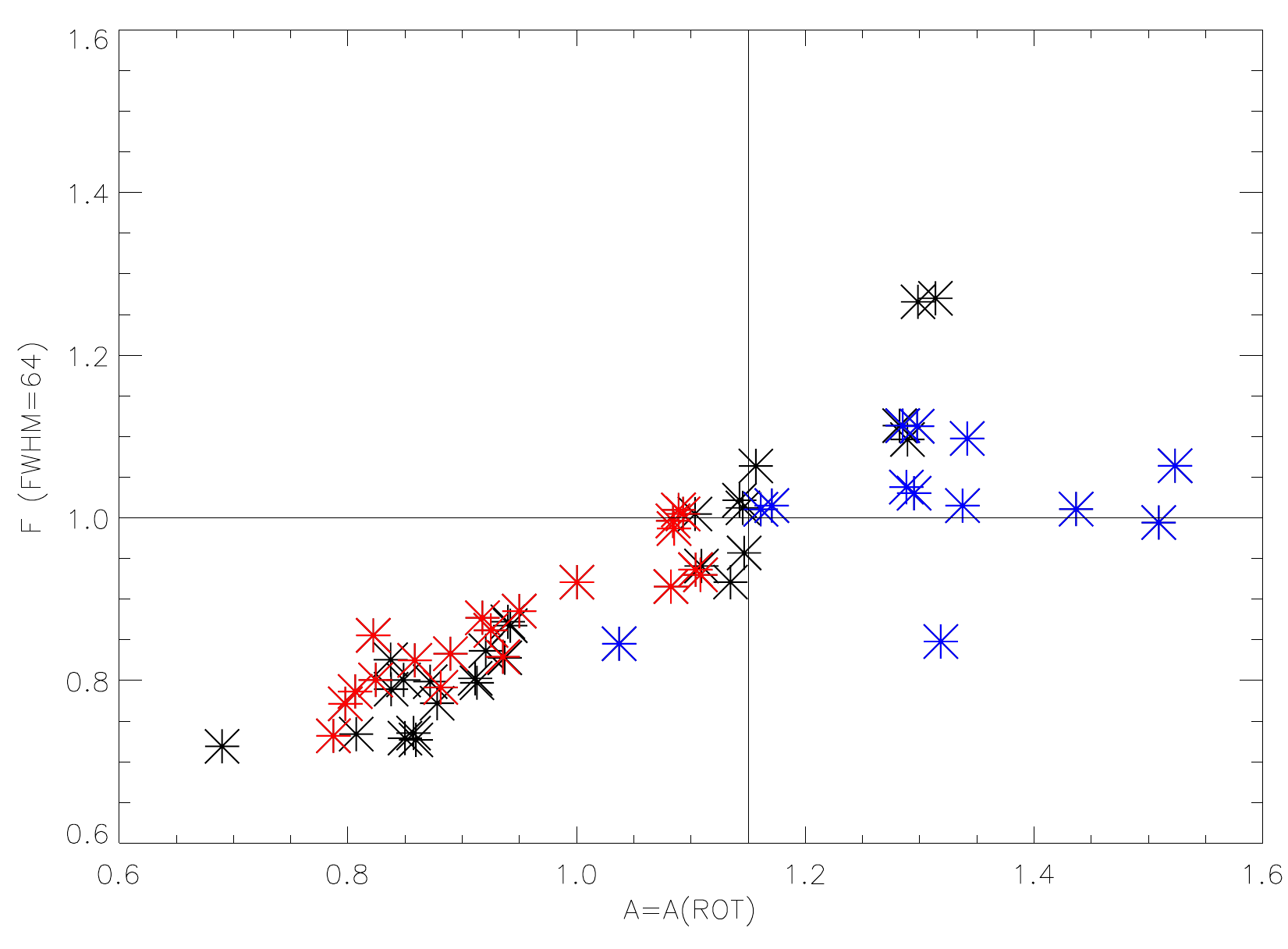}
\caption{Top panel: distribution of the {\it regular} (red) and {\it disturbed} (blue) clusters as a function of the asymmetry parameters: $A_{\rm rot}$ (top panels), $A_x$ (central panels), and $A_{\rm max}$ (bottom panels).  The sum is computed over a circular region of radius specified at the top of the columns. Lower panel:  distributions of fluctuation ($F_1$) and asymmetry ($A_{\rm rot}$) parameters for all the clusters. The values of horizontal and vertical lines are listed in Table~\ref{tab:slopes}.}
\label{fig:af}
\end{figure}

We assess the performances of the asymmetry parameters following the same procedure. The results for the entire analysis are reported in the bottom panel of the table (Table~\ref{tab:stat_f}). The distribution of $A$ for  \reg\, and \dist\, objects are shown in Fig.~\ref{fig:af} only in the cases of rotation (top panels), of flipping around the $x-$ axes (central panels), and of the maximum of all asymmetry parameters. The case $A_y$ is extremely close $A_x$  and, therefore, we chose to avoid its listing in Table~\ref{tab:stat_f}. The asymmetry parameter distinguishes effectively between the two classes of clusters, in particular when Eq.~\ref{eq:a} is derived including large radii. Selecting clusters with $A_{rot}<1.1$ at $R_{500}$ only one \dist\, cluster (5\% of the sample) contaminates the sample of otherwise \reg\, objects. The two distributions are generally broad and often present two major peaks (this is especially true for the regular systems). 
The most notably advantage of using the maximum of the three asymmetry parameters is visible in the bottom-right panel where the two distributions show clearly separated peaks with small dispersion (see also Table~\ref{tab:stat_f}). Excluding this case,
$A_{\rm max}$ does not significantly improve the classification. For this reason, in the rest of the paper we will focus only on $A_{\rm rot}$ computed at $R_{500}$, the version most used in literature.

Selecting the clusters by combining the constraints on both the asymmetry and the fluctuation parameters, $A_{\rm rot}$ (at $R_{500}$)$<1.15$ and $F_1 ({\rm at ~} R_{500})<1.0$, quite robustly gives \reg\, objects (only one clear contaminant\footnote{We call {\it contaminant} a cluster non part of the `mix' class and lying in the parameter space typical of the other morphological class.}) while objects with $A>1.15$ and $F>1.0$ are mostly \dist\, with no contaminants (see right panel of Fig.~\ref{fig:af}).
We conclude by noticing that the asymmetry parameter covers the same range of values as those derived by the Locuss collaboration \citep{zhang.etal.10,okabe.etal.10}.  On the other hand, we cannot compare our and their fluctuation parameters since 
$F$ is very sensitive to the area over which the sum of Eq.~1 extends and none of these papers specifies the maximum external radius used. For our simulated sample, adopting their formula we obtained $F$ values from 0 to 0.03 when we integrate over $R_{500}$ and from 0.03 to 0.14 when we stop at half of that radius.

\paragraph{X-ray SB concentration}

\begin{figure}[h!]
\centering
\includegraphics[width=0.45\textwidth]{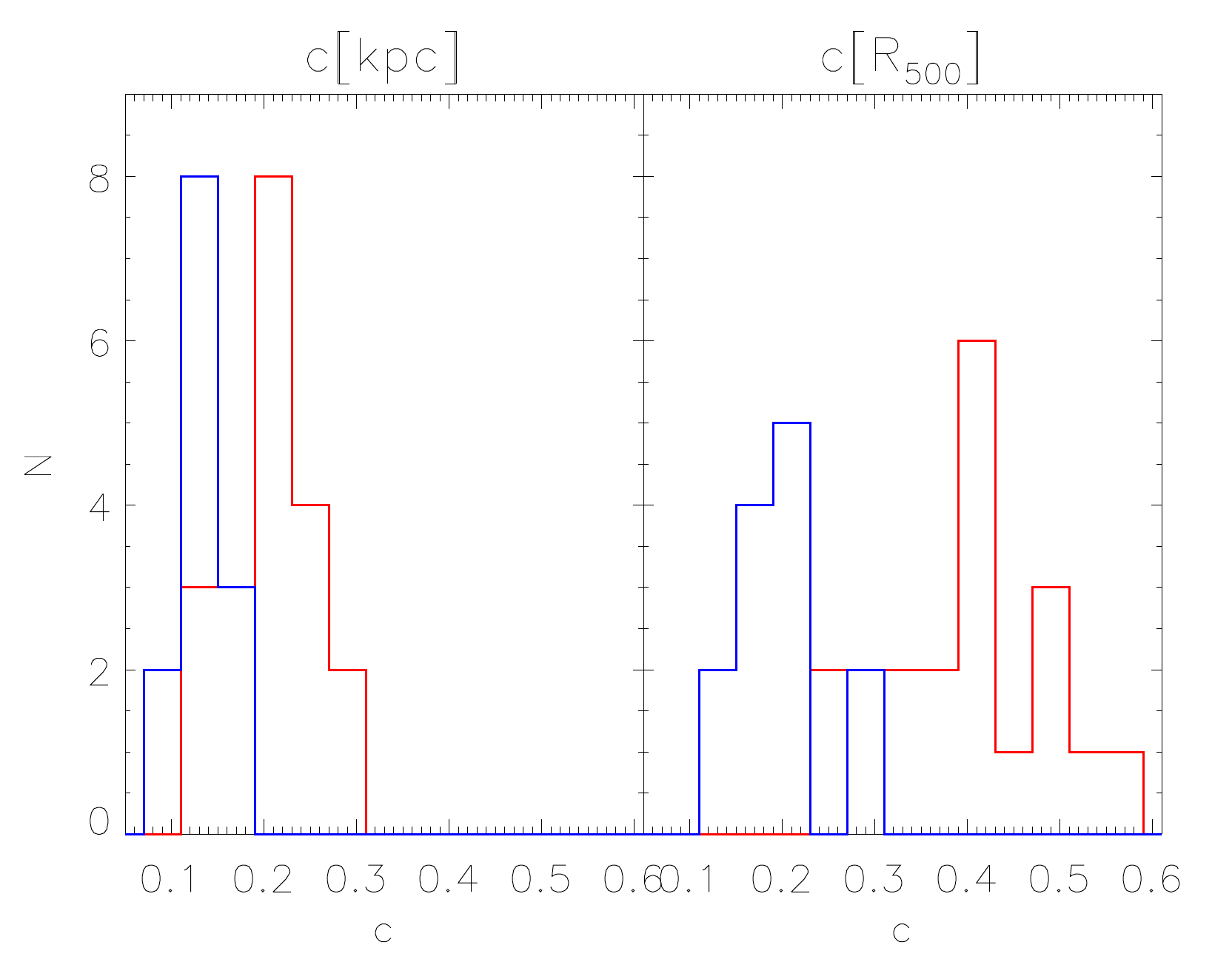}
\caption{Distribution of the {\it regular} (red) and {\it disturbed} (blue) clusters as a function of the X-ray surface brightness concentration $c_{\rm kpc}$ on the left and $c_{\rm R_{500}}$ on the right.}
\label{fig:ctot}
\end{figure}

The two definitions of X-ray surface-brightness concentrations ($c_{[kpc]}$ and $c_{[R_{500}]}$) are quite different since the $R_{500}$ values of our clusters range from $\sim$ 1 Mpc to $\sim$ 1.5Mpc (see Table 1 of R12 where the radii are reported in \hinv kpc). The results related to this parameter are reported in Table~\ref{tab:stat_cw} and shown in Fig.~\ref{fig:ctot}. When the measurements are performed using the physical radii the two distributions have very small dispersions and two clearly distinct peaks that, however, are as close as to produce an overlap of the tails of the distributions (left panel). The histograms significantly shift apart when the two radii used to compute the surface brightness are in units of $R_{500}$ (right panel). Both approaches are well suited to create subsamples of objects. In particular, in the first case, none of the objects with $c_{\rm kpc}<0.15$ ($c_{\rm kpc}> 0.2$) are \reg\, (\dist), in the second case a similar situation is presented for $c_{\rm R_{500}}<0.25$ ($c_{R_{500}}>0.3$). Even so, there is a significant difference between the numbers of objects that are located in between these boundaries: 4 \reg\, (20\% of the \reg\, sample) and 7 \dist\, objects ($\sim$ 50\% of the \dist\, sample) have $c_{\rm kpc}$ between 0.15 and 0.2, while only 2 \reg\, (10\%) and 2 \dist\, systems (15\%) have $c_{\rm R_{500}}$ between 0.25 and 0.3.  From Table~\ref{tab:stat_cw}, we notice that our overall distribution has the same median of the \chandra\, sample studied in \cite{cassano.etal.10}. 

\begin{table*}
\caption{X-ray surface brightness concentration, centroid shift, and third order power ratio parameter distributions (median, first and third quartiles) for the regular sample, all the clusters, and the disturbed sample. Last column refers to the Kuiper-two-sample-test probability.} 
\centering
\begin{tabular}{|c|ccc|ccc|ccc||c|}
\hline
$par$ &\multicolumn{3}{c|}{\reg} &\multicolumn{3}{c|}{all} &\multicolumn{3}{c||}{\dist} &  \\
 & $q^{\rm 1st}$ & med & $q^{\rm 3rd}$ & $q^{\rm 1st}$ & med & $q^{\rm 3rd}$  & $q^{\rm 1st}$ & med & $q^{\rm 3rd}$ &  $p$\\
\hline
& & & & & & & & & & \\

 $SB(0.2\times R_{500})/SB(R_{500})$   &   0.35 & 0.42  &  0.48  &  0.27  &   0.37   &  0.44 &   0.19 &  0.21 &  0.23 & 7 E-4 \\
 $SB(100 {\rm kpc})/SB(500 {\rm kpc})$  &   0.20 & 0.24  &  0.25  &  0.16  &   0.20    & 0.25 &   0.14 &  0.15 &  0.16  & 0.004 \\
& & & & & & & & & & \\
\hline
\hline
& & & & & & & & & & \\
${\rm log}_{10} (w)$                                   & -2.39  & -2.09  &  -1.72  & -2.09  &  -1.78 &   -1.49  & -1.68 &  -1.20 &  -1.10 & 0.020\\
& & & & & & & & & & \\
\hline
& & & & & & & & & & \\
$P_3 (R_{500})$                  &     -7.54    &   -7.21   &    -6.94  &      -7.13  &     -6.75  &     -6.15   &     -5.73  &     -5.33   &    -4.97  &    2 E-5 \\
$P_3 (0.5 \times R_{500})$ &     -7.32    &   -6.95    &   -6.48  &      -7.06   &    -6.52  &     -5.84    &    -5.79  &     -5.59    &   -5.34  &    0.001 \\
$P_3 (0.4 \times R_{500})$ &     -7.01    &   -6.64    &   -6.38  &      -6.77   &    -6.45  &     -5.93    &    -6.34  &     -5.63    &   -5.47  &    0.127 \\
$P_3 (0.3 \times R_{500})$ &     -6.86    &   -6.59    &   -6.38  &      -6.84   &    -6.48 &      -6.05    &    -6.61  &     -6.29    &   -5.77  &    0.168 \\
& & & & & & & & & & \\
$P_3$ (500 kpc)                  &    -7.51     &  -7.09      & -6.53 & -7.29     &  -6.53    &   -5.84  & -5.72     &  -5.32     &  -5.21  & 4E-5\\
$P_3$ (1000 kpc)                &   -7.65      &  -7.18      & -7.03 & -7.11      & -6.75     &  -6.17  & -5.82      & -5.26      & -5.00  & 8E-5 \\
& & & & & & & & & & \\
\hline
& & & & & & & & & &  \\
$P_{3, {\rm max}}$                        & -6.65   &    -6.45    &   -6.26  &  -6.41   &    -6.17  &      -5.60 &  -5.47   &    -5.13  &     -4.76 &  7 E-5\\
& & & & & & & & & & \\
\hline 
\end{tabular}
\label{tab:stat_cw}
\end{table*}

\paragraph{Hardness ratio indicators}

\begin{figure}[h]
\centering
\includegraphics[width=0.45\textwidth]{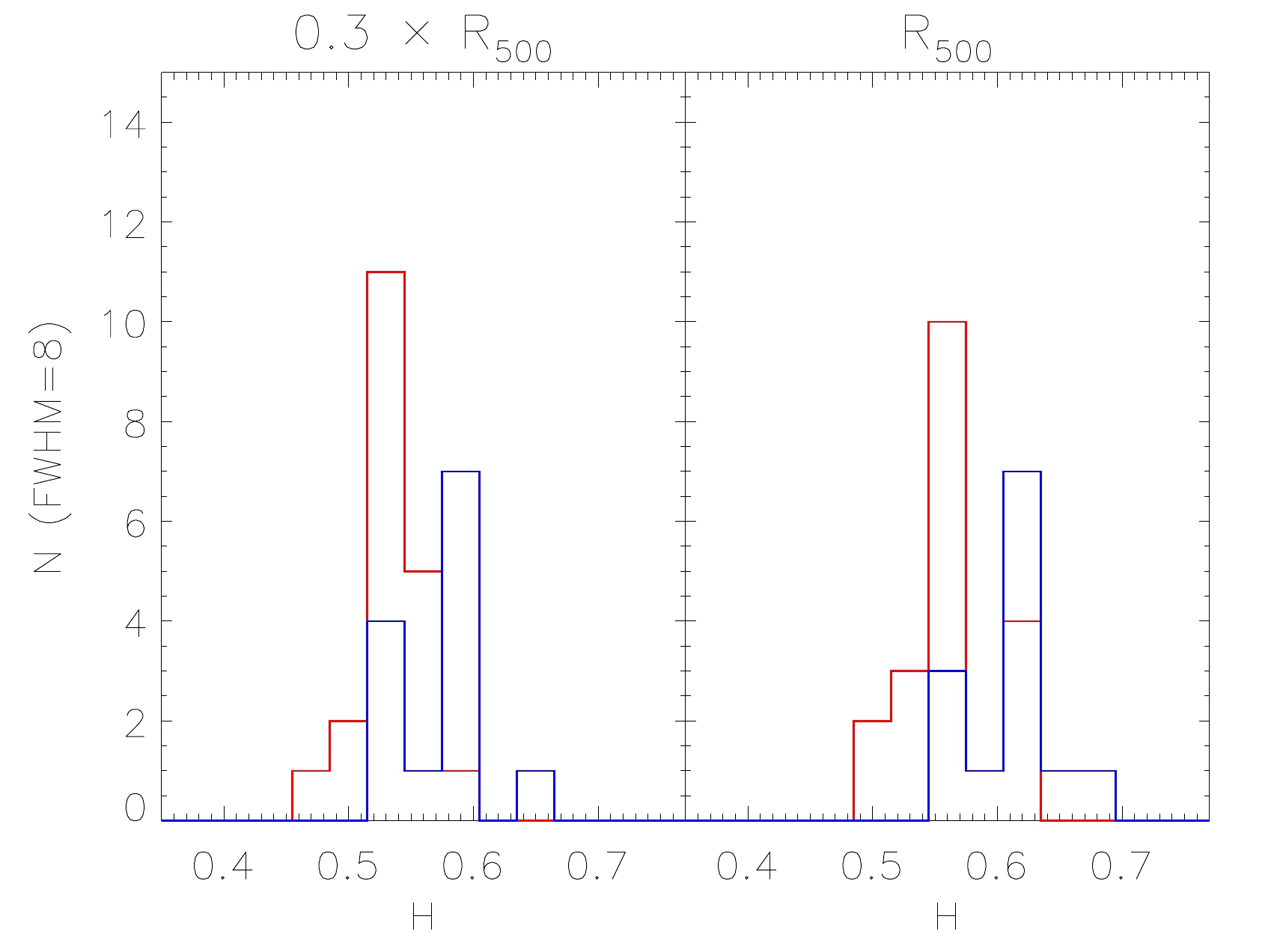}
\caption{Distribution of the {\it regular} (red) and {\it disturbed} (blue) clusters as a function of the fluctuation parameter, $H$. The smoothing image used in Eq.~\ref{eq:h} assume a Gaussian filter of FWHM equal $16^{\prime\prime}$.  Only results referring to circular regions of radius $0.3 \times$\rfive (left panel) and \rfive (right panel) are shown.}
\label{fig:hard}
\end{figure}

We  find almost no difference between the distributions of regular and disturbed clusters classified on the basis of the  two hardness ratio parameters, $H$, for all the smoothing lengths applied and for all radial range considered. The $p$ value of the Kuiper-two-samples test associated to the two $H$ distributions of \reg\, and \dist\, objects is always above 0.055 meaning that the this parameter is not effective in distinguishing the two morphological classes. In Fig.~\ref{fig:hard} we present two examples of $H$  (with $S$ extracted in the [0.3-1.5] keV band and $H$ in the [1.5-7.5] keV band) computed after smoothing the images with a Gaussian of FWHM equal to $16^{\prime\prime}$ and extending the sum to the smallest ($R<0.3\times R_{500}$) and to the largest ($R< R_{500}$) circular region. In these cases, $p=0.187$ and $p=0.057$, respectively. All other histograms (including those obtained from a different definition of $S$ and $H$) look very similar: the \dist -clusters are characterized by two peaks one of which is always at low $H$ values and, therefore, within the range typical of \reg\, objects. Due to the no-success of the hardness ratio indicator, we are not reporting the median and quartiles of the distributions and we will not consider it any further.

\paragraph{Centroid-shift}

\begin{figure}[h]
\centering
\includegraphics[width=0.45\textwidth]{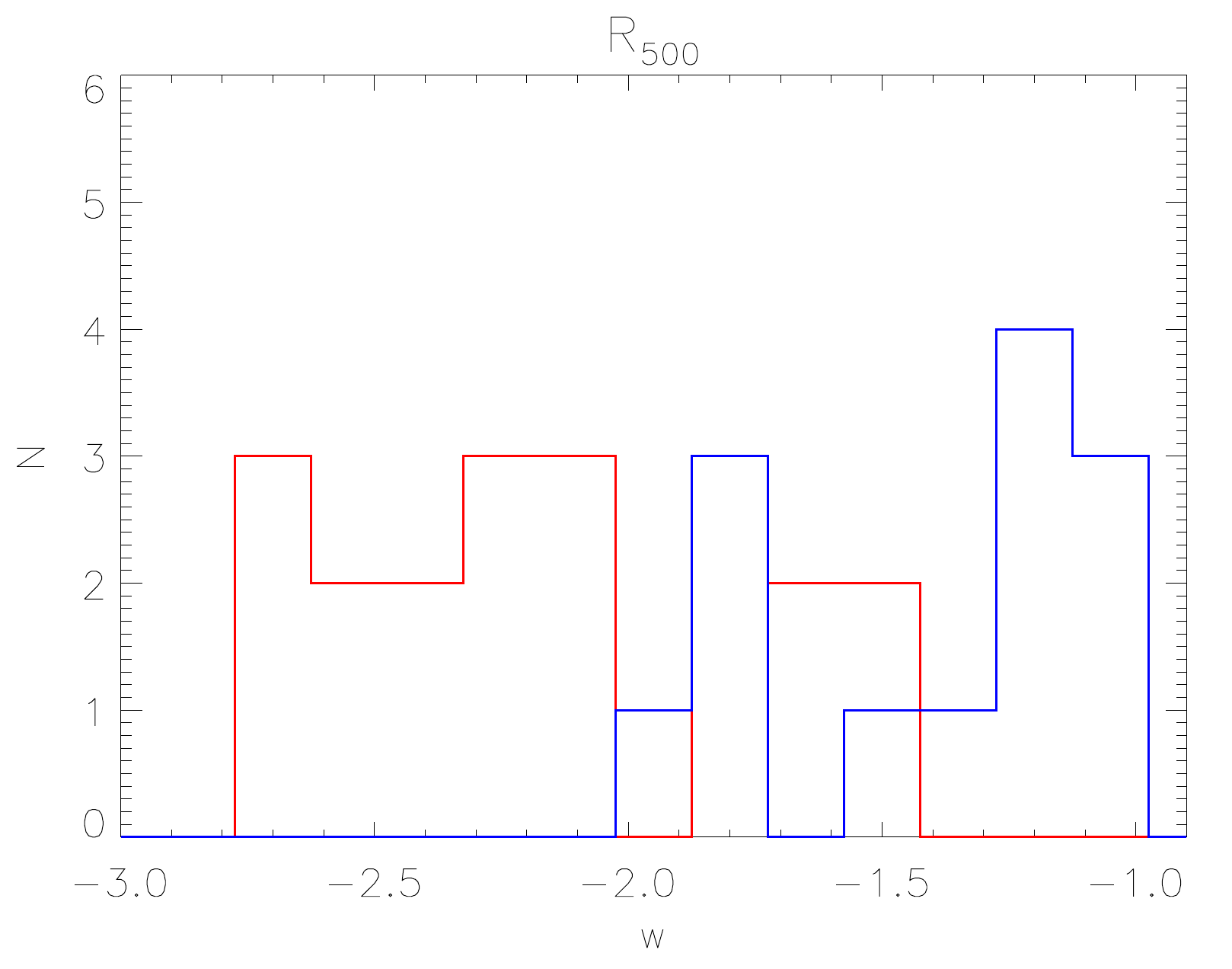}
\caption{Distribution of the {\it regular} (red) and {\it disturbed} (blue) clusters as a function of the decimal logarithm of the centroid-shift parameter, $\log_{10} (w)$.}
\label{fig:lw}
\end{figure}
The histograms of the centroid shift for the \reg\, and \dist\, clusters  present two well separated peaks  in case of $R_{\max}=$ \rfive (Fig.~\ref{fig:lw}). 
None of the objects having $w<0.015$ ($w>0.04$) is \dist\, (\reg). Nonetheless, between 0.015 and 0.04 the distributions overlap significantly. In this range, we find 30\% of the total distribution of the \dist\, systems and 35\% of the \reg\, ones. Therefore, even if the centroid shift is efficient in selecting systems that are part of only one category (the median are quite far apart) it also presents significant contamination.
The ranges of the centroid-shift obtained (from 0.002 to 0.12 ) agree well with the values from sample observed with \chandra,XMM-Newton, and ROSAT.

Subsequently, we allow $R_{\rm max}$ to vary from $R_{500}$ to $0.3 \times R_{500}$ with a step of $0.1 \times R_{500}$. We found that the Kuiper-two-sample probability increases and soon becomes unacceptable when the maximum radius decreases below $0.7 \times R_{500}$. 

\begin{figure}[h!]
\centering
\includegraphics[width=0.45\textwidth]{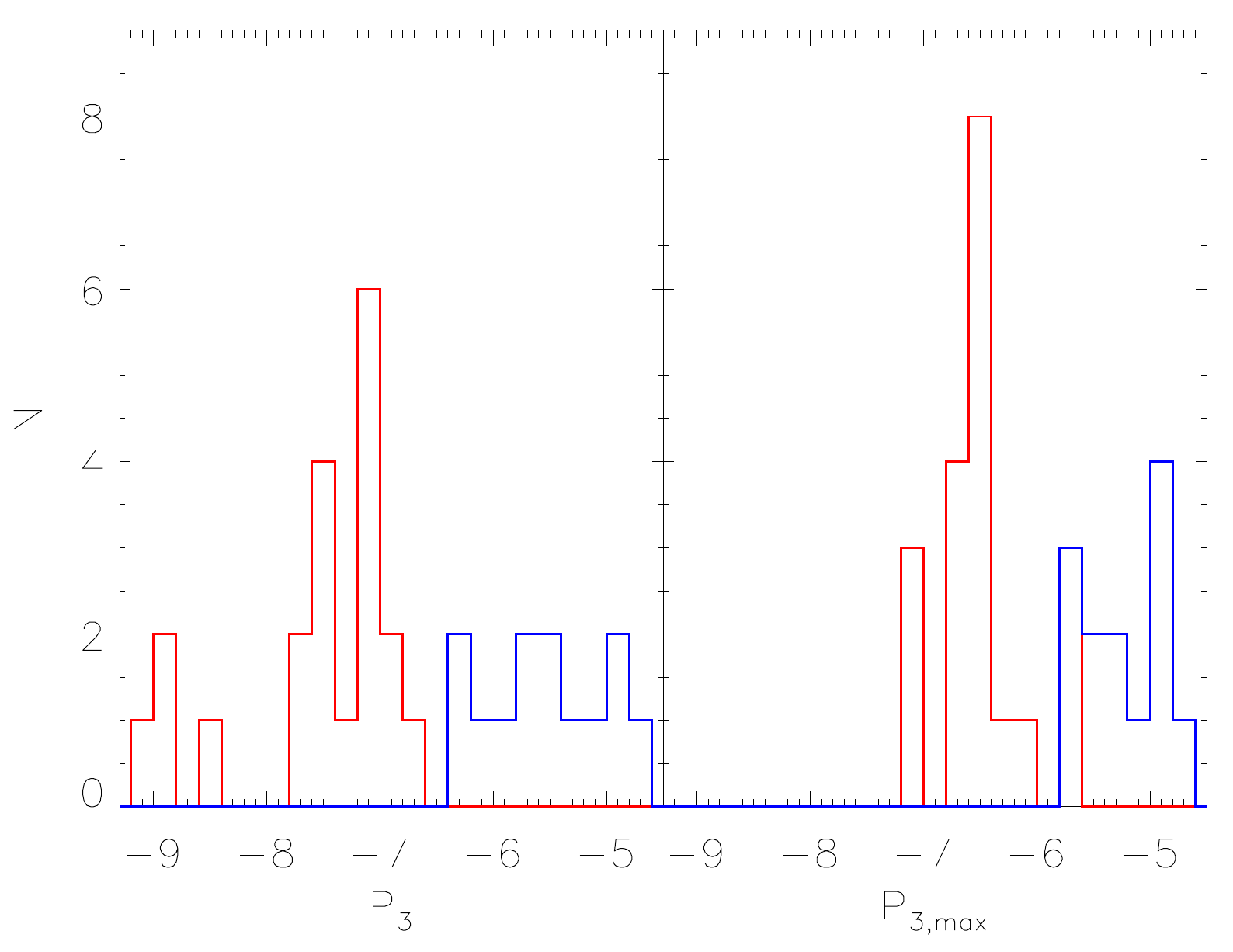}
\caption{Distribution of the {\it regular} (red) and {\it disturbed} (blue) clusters as a function of the third-order power ratio computed within $R_{500}$ (right panel) and maximum of third order power ratio (left panel).}
\label{fig:p3}
\end{figure}

\paragraph{Third-order power ratio and its maximum}
We  compute the third-order power ratio in eight different apertures: from $0.3 \times R_{500}$ to $R_{500}$ with step of $0.1 \times R_{500}$. For each of them we characterize the distributions of \reg\, and \dist\, systems and report the values of medians and quartiles in Table~\ref{tab:stat_cw}. The performances of the power ratio quite depend on the radius considered. The best results are obtained  for a large area (with $ R > 0.5 \times R_{500}$). Restricting the analysis to the central region, instead, impacts negatively on the efficiency of the method. 
Indeed, the $p$ value jumps from 0.001 (amply satisfying the requirement of $p<0.05$) at $0.5\times$ \rfive\, to 0.127 (meaning that the distributions of \reg\, and \dist\, systems are not clearly distinguishable) at lower radius.  On the left panel of Fig.~\ref{fig:p3} we show the result of the third order power ratio evaluated at $R_{500}$. The two distributions are significantly separated and no \reg\,  (\dist) objects are found with $P_3/P_0$ above $2 \times 10^{-7}$  (below $5\times 10^{-7}$). Similar behavior is found when the power ratio is measured within a fixed physical scale such as 500 or 1000 kpc (see Table.~\ref{tab:stat_cw}). Also in this case, we observed that measurements conducted over a larger area are more effective.

Subsequently, per each cluster we consider the maximum of the third-order power ratio profile (i.e. the maximum of the eight values computed) as suggested in \cite{weissmann.etal.13}. This case is shown in the right panel of Fig.~\ref{fig:p3} and reported at the bottom of Table~\ref{tab:stat_cw}. Per construction, all the numbers are shifted towards higher values causing an approach of the two medians. The two distributions are narrower and present higher peaks. No \reg\, objects are found with $P_{3, {\rm max}} > 3 \times 10^{-6}$ while only two disturbed systems are found below this limits (implying a 9\% of contamination).

\section{Pairs of parameters}

\begin{table}[ht!]
\caption{Coefficients of the linear fit to the points of various indicator pairs (1$^{\rm st}$ and 2$^{\rm nd}$ columns) and the Pearson correlation coefficient (3$^{\rm rd}$ column).} 
\centering
\begin{tabular}{|l|cc|c|}
\hline
$par$ &\multicolumn{2}{c|}{linear fit} & corr.   \\
pairs & $A$ & $B$ & coef.\\
 \hline
& & & \\
${\rm log_{10}}(P_3),A$  & 0.15  & 2.05       & 0.72\\
${\rm log_{10}}(P_3),c$  & -0.08 & -0.18     &  -0.73\\
${\rm log_{10}}(P_3),{\rm log_{10}}(w)$  &0.27 & -0.001     &  0.66 \\
$c,A$                               & -1.62 & 1.63      &  -0.87\\
${\rm log_{10}}(w),A$      & 0.31 & 1.61       &   0.66\\ 
${\rm log_{10}}(w),c$      & -0.19, &0.008    &   -0.77 \\
$A_{\rm rot},F$ & 0.55 & 0.33 &0.85\\
& &  &\\
\hline 
\end{tabular}
\label{tab:slopes}
\end{table}

\begin{table}[ht!]
\caption{Per each pair of indicator: values of the limit ($L$) dividing regular and disturbed objects  (vertical and horizontal in Fig.~\ref{fig:combo}) and the purity ($P$) and completeness ($C$) parameters.} 
\centering
\begin{tabular}{|l|cc|cc|cc|}
\hline
$par$ &\multicolumn{2}{c|}{ $L$} &\multicolumn{2}{c|}{\reg}& \multicolumn{2}{c|}{\dist}\\ 
pairs & ver.&hor. & $P_r$ &$C_r$ & $P_d$ & $C_d$\\
 \hline
& & & & &&\\
${\rm log_{10}}(P_3),A$  & -6.4    & 1.15  &  0.95 &0.95 &1.0 &  0.77\\
${\rm log_{10}}(P_3),c$  & -6.4    & 0.32   & 1.00 & 0.87 &0.9 &  0.45\\
${\rm log_{10}}(P_3),{\rm log_{10}}(w)$  &-6.4    & -1.4    &  0.95 & 0.95 & 1.0  & 0.45\\
$c,A$                               &0.32   & 1.15   &  1.00 & 0.87 & 1.0  & 0.72 \\
${\rm log_{10}}(w),A$      &-1.4   &1.15     & 0.93 & 1.00 & 1.0 & 0.36\\ 
${\rm log_{10}}(w),c$      & -1.4   & 0.32   &  0.97 & 0.87 & 1.0 & 0.45 \\
$A_{\rm rot},F$               &  1.15 & 1.00        & 0.89 & 0.89 & 1.0  &0.83\\
& & & & & &\\
\hline 
\end{tabular}
\label{tab:pure}
\end{table}

\begin{figure*}[h!]
\centering
\includegraphics[width=0.5\textwidth]{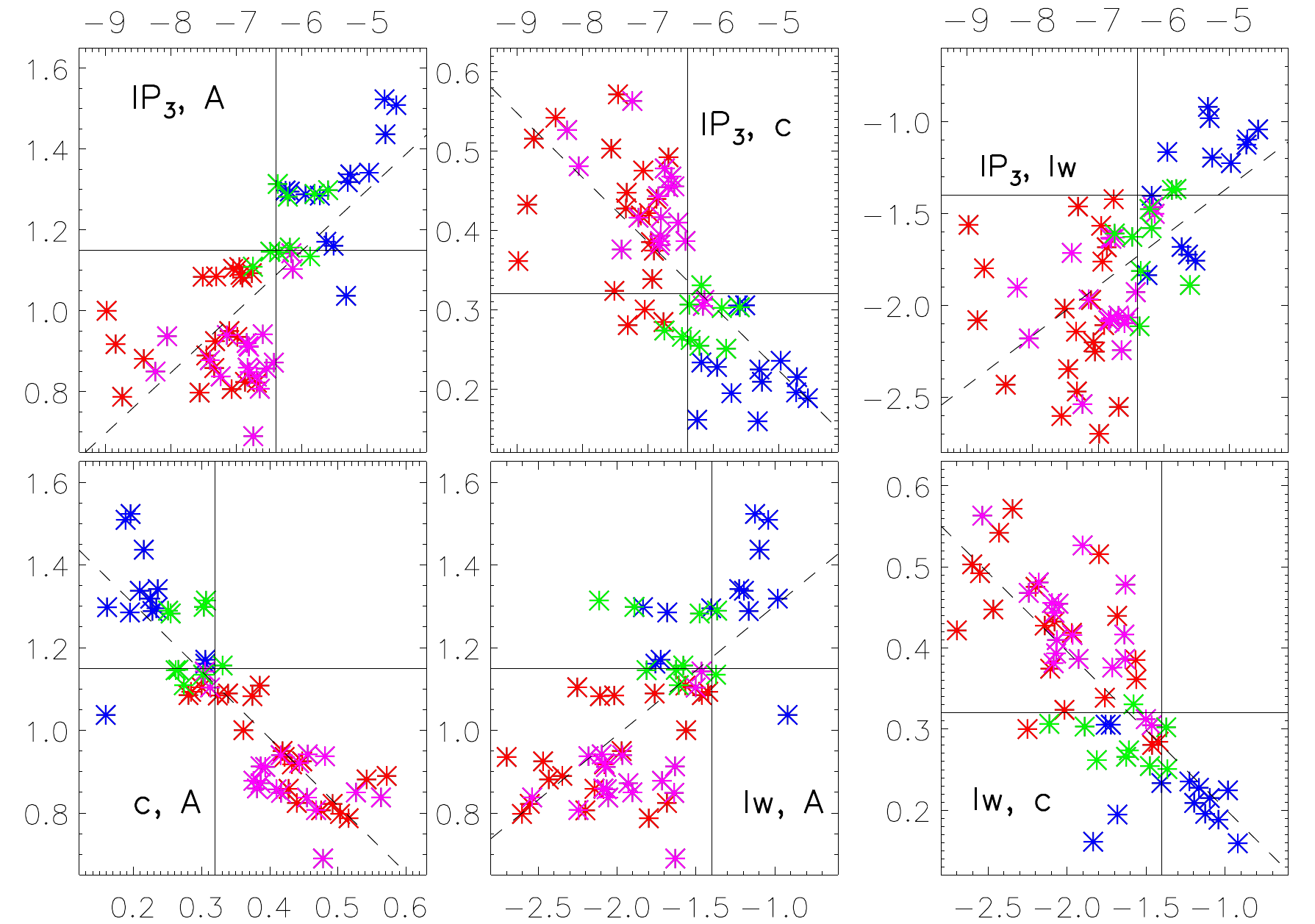}
\includegraphics[width=0.45\textwidth]{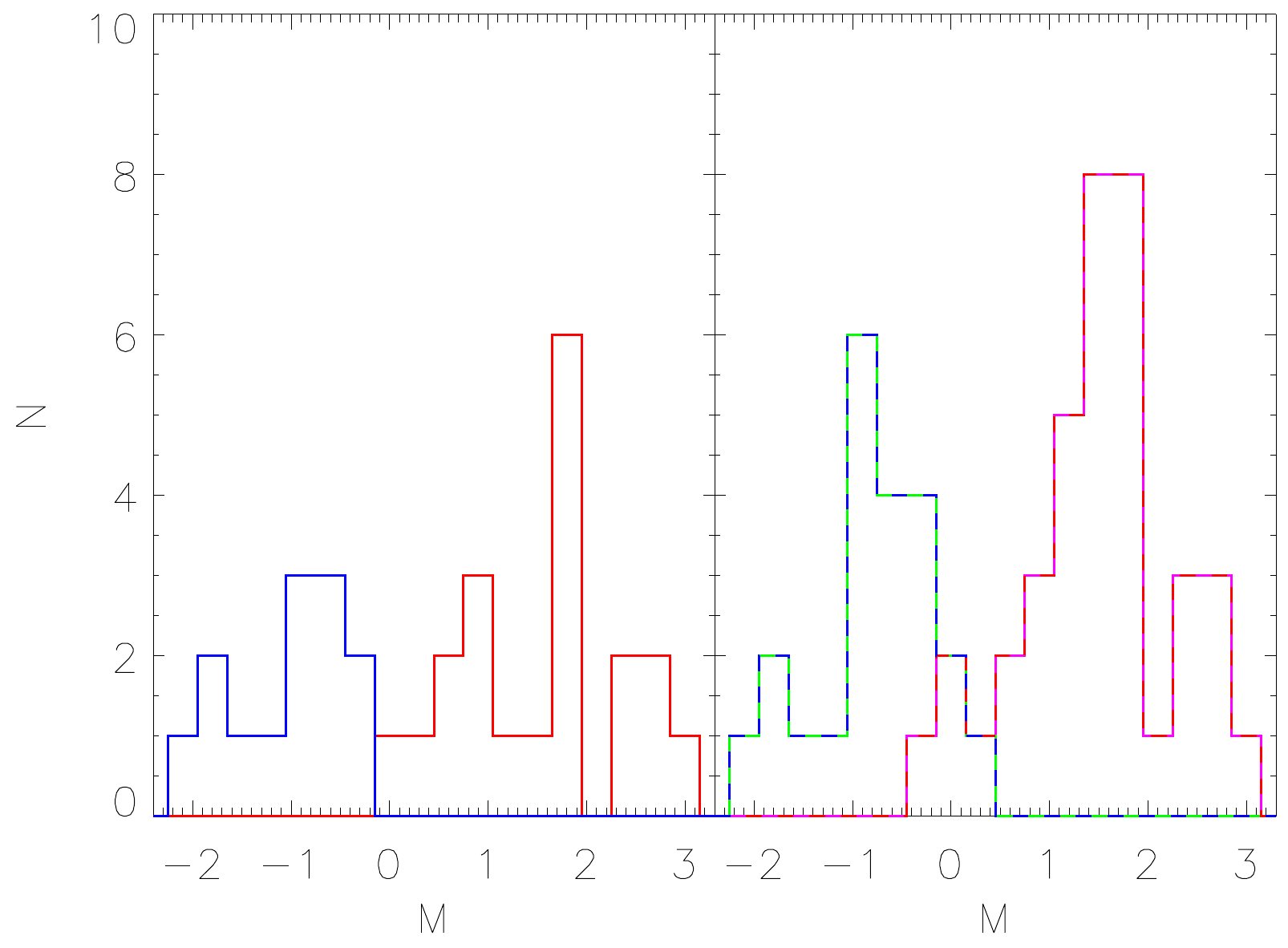}
\caption{Left panel: different combinations of the following parameters: decimal logarithm of the third order power ratio ($lP_3$), asymmetry parameter obtained by rotating the image ($A$), X-ray surface brightness concentration ($c$), and decimal logarithm of the centroid shift ($lw$). The corresponding ($x-y$) pair is written in each panel. Red, blue, green and magenta points denote \reg, \dist, {\it semi-regular}, and {\it semi-disturbed} clusters. The dashed lines are the best-linear fits to the points (the coefficients listed in Table~\ref{tab:slopes}). Right panel: distributions of the $M$ parameter (Eq.~\ref{eq:m}). Red and blue histograms refer only to \reg\, and \dist\, systems while red-magenta and blue-green distributions include {\it semi-regular} and {\it semi-disturbed} objects, respectively.}
\label{fig:combo}
\end{figure*}

In this Section, we study the combination of different pairs of indicators and test their robustness as we did for the asymmetry and the fluctuation parameters (Fig~\ref{fig:af}). All planes of the pairs studied are divided in four quadrants by the {\it limits}, $L$, listed in Table~\ref{tab:pure}. The parameters used are: the asymmetry parameter $A_{\rm rot},$, the decimal logarithm of the third order power ratio evaluated within $R_{500}$, the X-ray concentration $c_{\rm R_{500}}$, and the decimal logarithm of the centroid-shift. The other parameters (asymmetry computed by flipping the image around either a vertical or an horizontal axes, hardness ratios, third order power ratios within smaller circular regions, and the maximum of the third order power ratio) are not considered  here since their performances are either worse or similar to the selected indicators.
The results are shown in Fig.~\ref{fig:combo}. Per each combination, we also compute the linear fit and the Pearson correlation coefficient (Table~\ref{tab:slopes}). All pairs show a significant correlation degree (correlation coefficient always above 0.66). The pairs that correlate the least are those involving the centroid shift that for our sample  poorly accomplish the result of dividing the two samples and present a large scatter of values (see Fig.~\ref{fig:lw}). The pairs that, instead, correlate the most are those combining the asymmetry parameter with either the concentration or the fluctuation parameters. However, looking at Fig.~\ref{fig:af} and Fig.~\ref{fig:combo}, we can notice that several objects ($\sim 5$) clearly identified either as \reg\, or \dist\, are not located in the quadrant defining their morphological class.

To establish the best combination of parameters, we, therefore, calculate the `purity' and `completeness' of the \reg\, and \dist\, classes. To enlarge the statistics, we include the `mix' class spitting it into two other classes: {\it semi-regular} and {\it semi-disturbed} (see Appendix). In Fig.~\ref{fig:combo}, these two categories are shown by magenta and green points, respectively. 
We proceed then to evaluate the purity and completeness of the two extended classes. In the case of the \reg\, plus {\it semi-regular} systems (20+18 clusters) the two parameters are, respectively, defined as:
\beq
P_r=\frac{QN(\reg)}{QN(\dist+\reg)} \ \ \ \ {\rm and} \ \ \  C_r=\frac{QN(\reg)}{TN(\reg)}
\eeq
where $QN$ represents the number of objects in the quadrant and $TN$ the total number of objects. In similar way, we also compute $P_d$ and $C_d$ with respect to the \dist\, plus {\it semi-}\dist\, class (13+9 objects). The resulting values are reported in Table~\ref{tab:pure} together with the limits $L$ chosen per each parameter.

From this analysis, several interesting conclusions emerge. Purity of  both morphological classes is very high (always above 95\% except two cases) implying that there is very little contamination when two indicators are combined. The double boundaries, therefore, clearly enhance the power of the estimators in distinguishing between the two morphological classes. In most case, nevertheless, a good number of objects is lost, especially in the case of the \dist\, sample. Premise that this result might be influenced by the smaller number of \dist\, objects (the lost of four systems reduces $C_d$ to  $\sim$80\%), we notice that completeness is particularly poor whenever the centroid-shift parameter is involved (45\%, 36\%, and 45\% for the combination with  ${\rm log_{10}}(P_3/P_0)$, A, c, respectively). Taking into account all the parameters investigated the best combinations involve the asymmetry parameter and either the third order power ratio or the concentration parameter.

Encouraged by the improvement reached by joining two indicators, as final step, we defined a new parameter that combine a larger number of parameters. Considering the ones studied in this Section ($par =[log_{10}($\pthird$), log_{10}(w), c, A, F]$), $M$ is as follows:
\beq
M=\frac{1}{N_{par}}\sum_{par} A_{par} \times  \frac{par-L_{par}}{|q_{par}-m_{par}|}.
\label{eq:m}
\eeq 
$N_{par}$ is the number of parameters analyzed ($N_{par}=5$ in our case).
$A_{par}$ regulates the sign: it is considered equal to +1 only in the case of $par=c$, otherwise it is fixed equal to -1. In this fashion,  \reg\, systems should have a positive $M$. $L_{par}$ and $m_{par}$ are, respectively, the limits used in Fig.~\ref{fig:combo} (and listed in Table~\ref{tab:pure}) and the medians of the distribution of the parameters $par$  extended to the whole sample (Table~\ref{tab:stat_f} and Table~\ref{tab:stat_cw}). Finally, $q_{par}$ is either the first or the third quartile depending if  the parameter of the specific cluster is smaller or larger than its median, $m_p$. In other words, $M$ is the sum of all differences of the parameters from their limits normalized by the distance from quartiles to the medians. In the left panel of Fig.~\ref{fig:combo}, we are showing the histograms of $M$. Red and blue distributions represent \reg\, and \dist\, objects while red-magenta and blu-green ones include also {\it semi-regular} and {\it semi-disturbed} systems. The separation between the two more restrict morphological classes is clear, and the overlap for the extended classes is limited to three objects, all close to the 0 boundary. All objects with $M>0.05 (<-0.05)$ are \reg\, (\dist).  As evident from the image, the Kuiper-two-sample-test probability, in both cases, is extremely low ($p=$ 8 E-5 and $p=$1 E-8, respectively). One of the most remarkable feature is that the majority of the objects in both morphological classes are located in regions significantly apart from 0. This is especially true for the \reg\, clusters whose distribution's peak is located around 2, meaning that most of the clusters have more than one parameter significantly beyond the quartile of its distribution. 
 From comparing the distribution plots in Figures~1 through 5 with the quartiles values listed in Table~1 and 2, we evince that the $M$ parameter is the only one presenting this characteristics. For all the other parameters, indeed, the peak of the \reg\, clusters' distribution is around to the value of the quartile ( M $\sim$ 1).
Since the peaks of the two distributions are far apart from the boundary and greatly apart from each other, $M$ is the strongest measure of the X-ray morphology. We further compute $M$ assuming $N_{par}$ equal to 4 (removing the centroid shift) and to 3 (removing also the fluctuation parameter). The distributions in both cases are very similar to the ones plotted in Fig.~\ref{fig:combo} (results not shown). However, we noticed that the two morphological classes have higher peaks when we use all five indicators. The only advantage we witness by excluding the centroid-shift parameter is that the peak of the \dist\, class moves farther away towards more negative numbers.  The last but probably most important feature of the $M$ parameter is that it is equivalently strong in distinguishing the two morphological classes even when in Eq~11 we use the median of each parameters instead of the limit $L_{par}$. Also in this context the distribution of the total sample is clearly binomial (such as the distribution shown in the right panel of Figure~7). This property is not present in any of the individual parameter for which none of their median is equivalent to the boundary, $L_{par}$, individuated in this work. Since $M$ efficacy does not depend on a {\it a priori} knowledge of the boundaries, it is the most accessible tool to use in future surveys (provided that the objects observed are numerous and not biased towards a particular morphological class).


\section{Conclusions}

Analyzing 60 synthetic \chandra\, images, we tested several morphological parameters commonly used in the literature to distinguish between \reg \, and \dist\, systems. 
The morphology of each image has been rated 9 times (three grades assigned by each of the three observers) with a grade from 0 to 3, where 0 indicates extremely regular objects and 3 very disturbed ones. All systems with an average grade $\leq 1.5$ or $>2.5$ were declared as \reg\, or \dist, respectively. In the final count the two classes have 20 and 13 objects. We used this definition to evaluate the capabilities of the estimators to identify systems clearly regular or clearly disturbed. Subsequently, we studied various combination of parameters. For this goal, we extend the analysis to the whole sample to enlarge the statistics and to be able to compute the purity and completeness parameter. We proceeded to sub-divide the `mix' class into {\it semi-regular} and {\it semi-disturbed} objects (see Appendix) fixing the diving grade equal to 2.  We decided to test the morphological parameters against the visual classification because our intent is to search for the best parameter that can be as accurate as a trained eye. However, {\it a posteriori}, we checked also at the mass accretion history of our objects and found that our \reg\, and {\it semi-regular} clusters have accreted on average only 10\% of their mass at $r_{200}$ in the last megayear against the 20\% of the most disturbed systems. These numbers grow, respectively, to 30\% and 50\% with respect to a 1.7 Myr time lapse. Looking at the mass accretion history of each individual cluster we found that all our \reg\, systems are growing their mass slowly with the only exception of CL20. This system experienced an almost simultaneous merger by three different structures that, however, are still outside the Chandra field of view. As a confirmation all the morphological parameters investigated returned values typical of a {\it regular} object.

For each estimator, we showed the distribution of the two extreme morphological classes and reported the median, 1$^{\rm st}$ and 3$^{\rm rd}$ quartile. We discussed the level of performance of each indicators by listing the percent of {\it contaminant} systems (defined as  clusters of a particular morphological class lying in the parameter range typical of the other class). Finally, we analyze the advantages of considering pairs of indicators over a single one. Following this logic, we also define a new parameter, $M$, that adds the performances of five indicators. This new parameter is the strongest measure to discriminate between morphological classes among those tested in this paper. 

Before summarizing our results, let us remind that the values of all the parameters computed using our synthetic catalogue agree with those derived from cluster samples based on observations apart the X-ray surface-brightness concentration parameter. Indeed, while the median value of this parameter is coincident with that found in \cite{cassano.etal.10}, our concentrations reaches at maximum the value of  0.4 while those from the observational sample extend to $c=0.7$. We advance two plausible explanations for this difference: $i)$  our clusters are all very massive and $ii)$ the majority of them lie in a dense environment (R12). The first fact leads to a lower concentration of mass (and, thus, light) at the center of the objects with respect to smaller groups \citep{nfw96}. The second situation implies that out systems have a large amount of mass in their outskirts since they are continually accreting from the surrounding filaments (even if not through major or significant mergers). We exclude that the concentration discrepancy  is caused by some numerical characteristics or by the physics adopted in the simulations. In fact, radiative simulations with no active and powerful feedback (such as ours) are predicted to produce brighter and denser cores \citep{borgani&kravtsov} with respect to real clusters.  Therefore if any contrast in concentration parameter is present due to the overcooling problem affecting simulations, this is expected to be on the other direction.
With the exclusion of this small difference, all other parameters agree with previous theoretical and observational measurements.  On that account, we can conclude that after the extraction of the central part  current complex theoretical models are not only able to  fairly well reproduce standard scaling relations \citep[e.g.]{fabjan.etal.10,fabjan.etal.11,puchwein.etal.08, battaglia.etal.12}, the main properties profiles such as density, temperature, pressure, and X-ray surface brightness \citep[e.g.][]{nagai.etal.07,fabjan.etal.10,pratt.etal.07,mroczkowski.etal.09,eckert.etal.13,arnaud.etal.10} but also the overall appearance of real clusters.

 This strengthen our analysis whose results can be directly applied to up-coming surveys. Concluding, we found that:
\begin{itemize}
\item the fluctuation parameter performs better if computed considering the entire region within $R_{500}$ and a broader-smoothed (2$^{\prime}$) image. Also in this occasion, however, three \dist\, halos (23\% of the \dist\, sample) fall within the \reg-system range. The efficiency of this parameter is improved when associated with the asymmetry one, assuming $F=1$ and $A=1.15$ as dividing points within the two morphological classes;
\item the asymmetry parameter is one of the clearest indicator of morphology. The distributions of the \reg\, and \dist\, objects are well separated in case of $A_{\rm rot}$ calculated within the $R_{500}$ region. There is no improvement in adopting variation of the asymmetry parameter (obtained, for example, by flipping the image instead of rotating it) nor in taking the maximum among different asymmetry parameters. This parameter is specially effective when it is combined with the X-ray concentration or the third order power ratio;
\item the X-ray surface brightness concentration parameter competently divides the two morphological classes especially when computed using radii expressed in units of $R_{500}$. The contamination is minimal, the two distributions are well separated as are their peaks;
\item the hardness ratio is the only parameter among those investigated that is not able to discern among \reg\, or \dist\, objects. This might depend on the fact that  both pre-merger and post-merger clusters belong to the last category. The thermalization states of this class, therefore, can be quite diverse;
\item the centroid-shift estimated in our analysis deludes our expectations in terms of characterizing the \dist\, clusters. We believe this is due to characteristics of our sample that present two highly complex clusters with an extended X-ray emission symmetric to the center (CL17 in both directions) and another object with a large substructure but still in the outskirts (CL15 both directions). All \reg\, systems have a centroid-shift below 0.04 (-1.4 in decimal logarithm), however, also several contaminants are present, including CL 15 and CL17. Improvements in purity and completeness of the \reg\, sample can be registered if this parameter is associated to others. However, even with this stratagem the completeness of the \dist\, sample remains poor (below 50\%). For the large contamination present at middle values of this parameter, we consider the best approach to identify two separate limits. For example, in our sample, all objects with $w>0.04$ are \dist\, while the majority of systems with $w<0.01$ is \reg\, (with the exception of one disturbed system having $w=0.008$);
\item the third order power ratio was estimated using various apertures. The most remarkable result was achieved when the parameter was computed within $R_{500}$ or a fixed physical scale of 1000 kpc. In both cases the histograms related to the two distributions are completely separated. The calculus of the power ratio limited to the very center of the cluster is inadequate for our purpose. The adoption of the peak of the power ratio profile does not improve the result substantially even if it has the gain of narrowing the distributions and increasing their peaks. The advantages of this last parameter can be enhanced if in the sample there are various systems with major substructures close to the center of the cluster. This, however, was not the case for the objects considered in this work.
\item In general terms, all morphological parameters show a strong correlation. For this reason, combining two or more parameters is a more efficient way to differentiate the morphological classes.
\item  We defined a new morphological parameter called $M$. This is the most powerful parameter since it combines the strengths of all the individual parameters (it incorporates detection of active dynamical history, captured by $w$, complexity and presence of substructures, typically measured with $A$, $F$ and $P_3$). With respect to the individual parameter, $M$ has two major advantages: it clearly has a bimodal distribution with far apart peaks and it is successful in distinguishing between the two morphological classes also when we use the median as reference in Eq~1, i.e. without any {\it apriori} condition on the boundary to be used.  Studying various combination of the parameter $M$ we can further determine the performances of the parameters involved. Indeed, by simply comparing the distributions of $M$ with and without the centroid-shift parameter we are able to  state that this indicator is not suitable to uniquely categorize the \dist\, class in our sample.
\end{itemize}

Finally, it is always the case that measurements conducted within large scales (either $R_{500}$ or 1000 kpc) perform more efficiently. This appear the best situation from a theoretical point of view, however, real clusters suffer from problematics that we have not considered in this paper. We refer to \cite{weissmann.etal.13} for suggestions on the best approach to take in order to reduce the effect of Poisson noise and X-ray background.
 

\begin{figure*}[h!]
\centering
\includegraphics[width=0.8\textwidth]{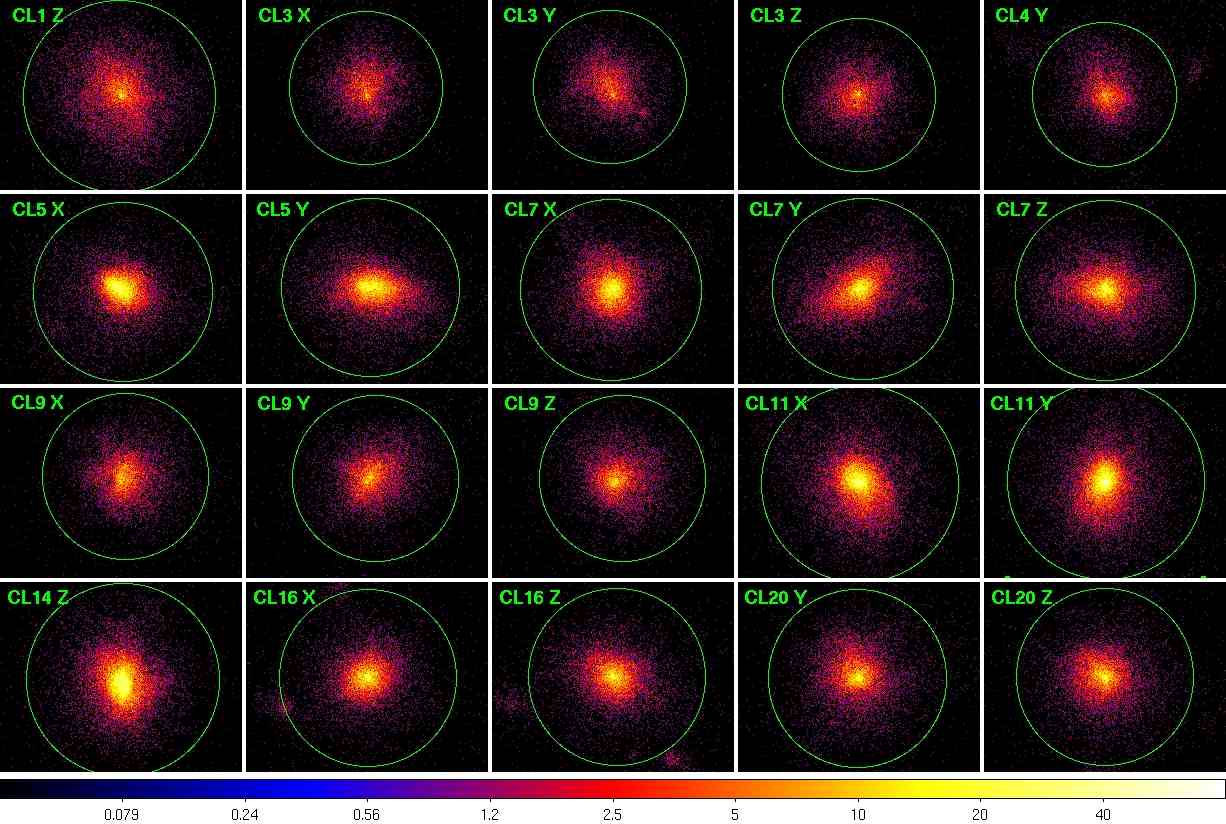}
\caption{X-ray soft images (0.7-2) keV of the cluster visually classified as \reg. The green circle indicating $R_{500}$ has been center on the X-ray centroid.}
\label{fig:reg}
\end{figure*}

\begin{figure*}[h!]
\centering
\includegraphics[width=0.8\textwidth]{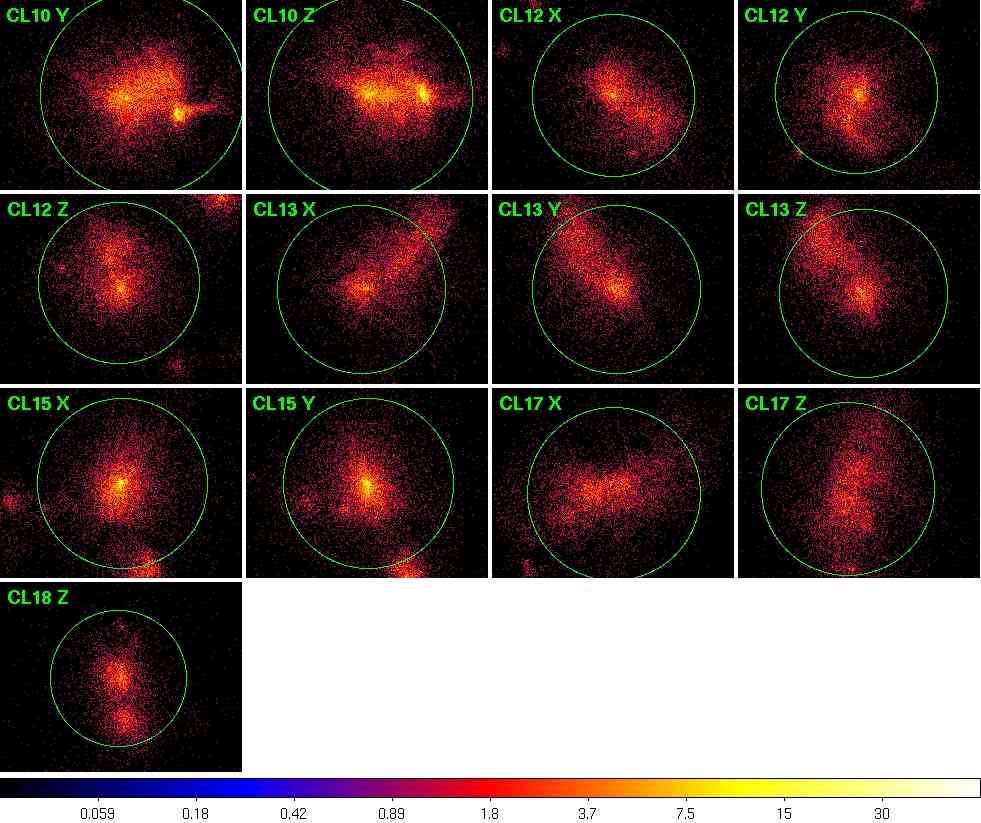}
\caption{X-ray soft images (0.7-2) keV of the cluster visually classified as \dist. The green circle indicating $R_{500}$ has been center on the X-ray centroid.}
\label{fig:dist}
\end{figure*}

\begin{figure*}[ht!]
\centering
\includegraphics[width=0.65\textwidth]{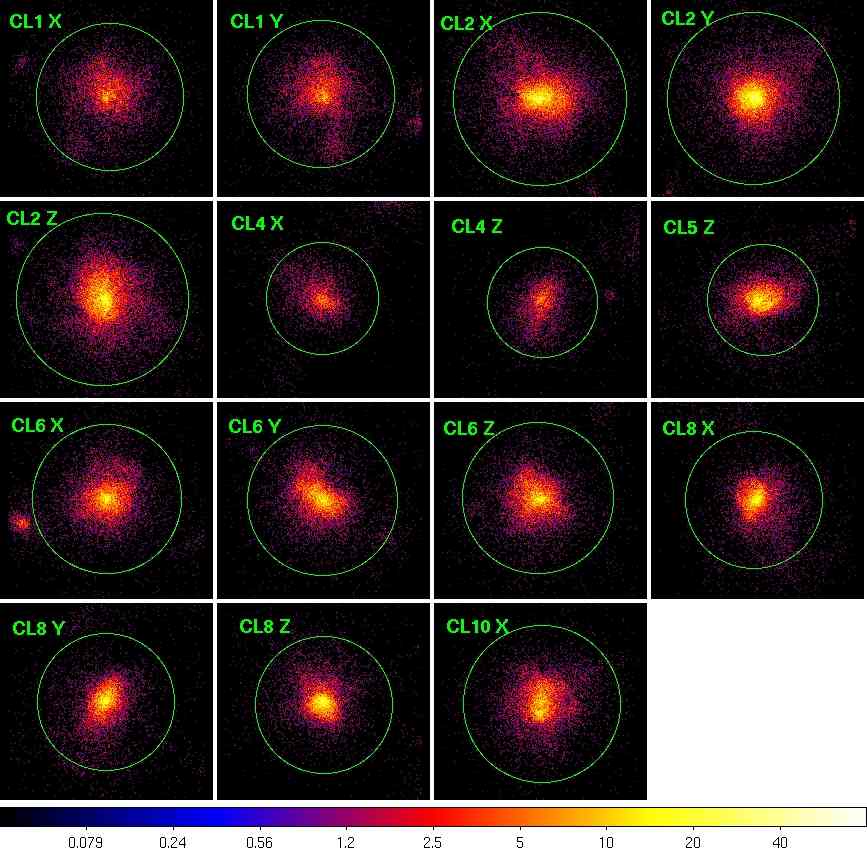} \\
\includegraphics[width=0.65\textwidth]{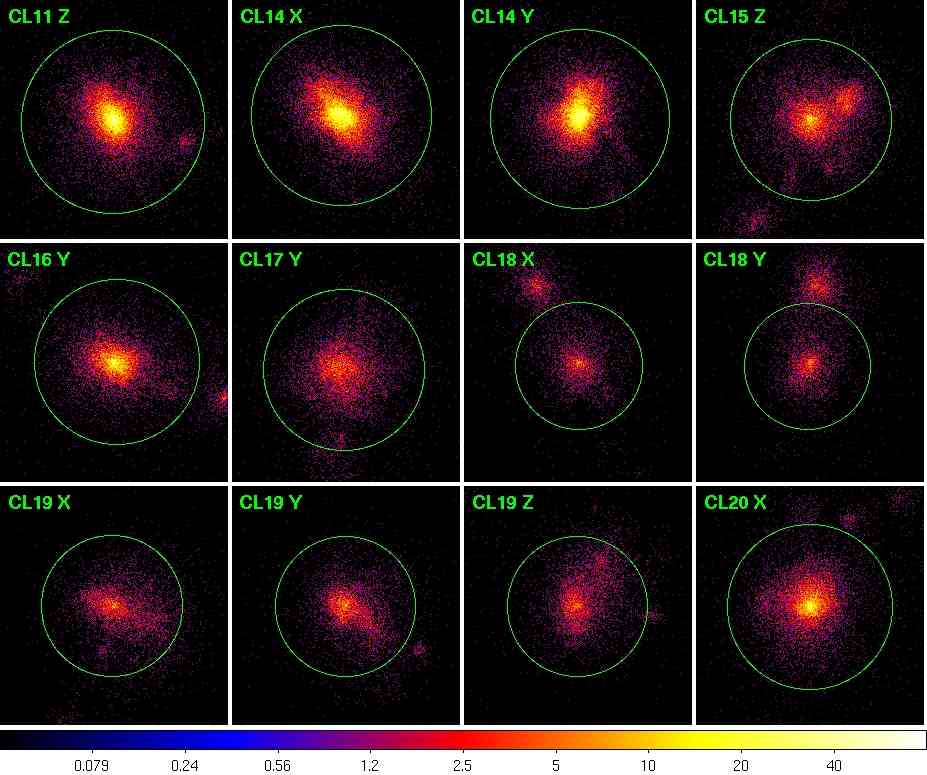}
\caption{X-ray soft images (0.7-2) keV of the cluster visually classified as mix. The green circle indicating $R_{500}$ has been center on the X-ray centroid.}
\label{fig:mix}
\end{figure*}

\bigskip
\begin{center}
{\bf Appendix}
\end{center}

In Fig.~\ref{fig:reg} and Fig.~\ref{fig:dist} we show the samples of \reg\, and \dist\, systems\footnote{Large resolution images are available  following the link to the synthetic cluster images at http://www-personal.umich.edu/$\sim$rasia}. The green circles in the images are centered in the X-ray centroid used for rotating and flipping the images to compute the asymmetry parameters. Their radii correspond to $R_{500}$ (see Table.~1 of R12). The cluster ID and the direction of the line of sight reported on the top-left corner follow the same nomenclature of R12. The mix class (Fig.~\ref{fig:mix}) has been divided into {\it semi-regular} (including CL2 X--Y--Z, CL 4 X--Z, CL5 Z, CL6 X--Y--Z, CL8 X--Y--Z, CL10 X, CL11 Z, CL14 X--Y, CL16 Y, CL20 X) and {\it semi-disturbed}  (including CL1 X--Y, CL 15 Z, CL17 Y, CL18 X--Y--Z, CL19 X--Y--Z) objects. 
 Notice that for almost all the objects, it is the case that the three images obtained by varying the line of sight projections belong to the same extended morphological class. The only exceptions are two clusters: CL1 and CL10. In Table~\ref{tab:all}, we list the main morphological parameters for the entire set of clusters.

\begin{table*}
\caption{List of parameter per each cluster evaluated with respect to $R_{500}$} 
\centering
\begin{tabular}{|c|ccccc|c|c|ccccc|}
\hline
CL & $F_1$ & $A_1$ & $c$ & $w$ & $P_3$&&CL & $F_1$ & $A_1$ & $c$ & $w$ & $P_3$  \\
\hline
& & & & & & & & & & & & \\
CL1 X &0.96 & 1.15 & 3.30e-07 &  0.27 & 0.024 &&CL11 X &0.77 & 0.80 & 2.75e-08 &  0.50 & 0.002 \\
CL1 Y &0.94 & 1.11 & 1.78e-07 &  0.27 & 0.024 &&CL11 Y &0.79 & 0.88 & 3.88e-09 &  0.54 & 0.004\\
CL1 Z &0.94 & 1.10 & 8.93e-08 &  0.30 & 0.006 && CL11 Z &0.79 & 0.84 & 5.76e-08 &  0.56 & 0.003\\
CL2 X &0.73 & 0.86 & 2.88e-07 &  0.41 & 0.009 && CL12 X &1.01 & 1.34 & 5.48e-06 &  0.21 & 0.064 \\
CL2 Y &0.73 & 0.81 & 2.28e-07 &  0.47 & 0.006 &&CL12 Y &1.03 & 1.29 & 6.54e-07 &  0.23 & 0.039 \\
CL2 Z &0.73 & 0.86 & 1.49e-07 &  0.38 & 0.008 &&CL12 Z &1.04 & 1.29 & 1.13e-06 &  0.23 & 0.068\\
CL3 X &0.99 & 1.08 & 4.92e-08 &  0.28 & 0.034 && CL13 X &1.06 & 1.52 & 1.82e-05 &  0.20 & 0.075\\
CL3 Y &1.00 & 1.09 & 1.72e-07 &  0.28 & 0.038 &&CL13 Y &1.01 & 1.44 & 1.89e-05 &  0.21 & 0.080\\
CL3 Z &1.00 & 1.08 & 3.07e-08 &  0.32 & 0.010 &&CL13 Z &0.99 & 1.51 & 2.75e-05 &  0.19 & 0.091\\
CL4 X &1.02 & 1.14 & 6.90e-07 &  0.30 & 0.034 &&CL14 X &0.72 & 0.69 & 1.81e-07 &  0.48 & 0.023\\
CL4 Y &1.01 & 1.09 & 1.16e-07 &  0.34 & 0.017 &&CL14 Y &0.73 & 0.85 & 5.82e-09 &  0.53 & 0.012 \\
CL4 Z &1.00 & 1.10 & 7.26e-07 &  0.31 & 0.031 &&CL14 Z &0.73 & 0.79 & 1.81e-09 &  0.52 & 0.016\\
CL5 X &0.83 & 0.89 & 3.48e-08 &  0.57 & 0.004 &&CL15 X &1.01 & 1.16 & 3.04e-06 &  0.31 & 0.017\\
CL5 Y &0.80 & 0.82 & 1.35e-07 &  0.44 & 0.021 &&CL15 Y &1.01 & 1.17 & 2.35e-06 &  0.30 & 0.019 \\
CL5 Z &0.80 & 0.85 & 1.58e-07 &  0.42 & 0.023 &&CL15 Z &0.92 & 1.13 & 1.33e-06 &  0.30 & 0.042\\
CL6 X &0.80 & 0.87 & 3.73e-07 &  0.39 & 0.012 &&CL16 X &0.85 & 0.82 & 2.06e-07 &  0.49 & 0.003 \\
CL6 Y &0.80 & 0.91 & 1.53e-07 &  0.39 & 0.023 &&CL16 Y &0.87 & 0.94 & 2.53e-07 &  0.46 & 0.008 \\
CL6 Z &0.77 & 0.88 & 3.95e-08 &  0.37 & 0.019 &&CL16 Z &0.86 & 0.92 & 4.73e-08 &  0.45 & 0.003\\
CL7 X &0.79 & 0.81 & 8.48e-08 &  0.48 & 0.006 &&CL17 X &1.11 & 1.28 & 1.88e-06 &  0.19 & 0.021\\
CL7 Y &0.83 & 0.93 & 1.03e-07 &  0.42 & 0.002 &&CL17 Y &1.01 & 1.14 & 4.46e-07 &  0.26 & 0.015 \\
CL7 Z &0.82 & 0.86 & 4.60e-08 &  0.43 & 0.007 &&CL17 Z &1.11 & 1.30 & 5.64e-07 &  0.16 & 0.015\\
CL8 X &0.84 & 0.92 & 1.43e-07 &  0.44 & 0.008 &&CL18 X &1.27 & 1.31 & 4.28e-07 &  0.31 & 0.008\\
CL8 Y &0.83 & 0.84 & 2.04e-07 &  0.45 & 0.009 &&CL18 Y &1.27 & 1.30 & 2.51e-06 &  0.30 & 0.013\\
CL8 Z &0.83 & 0.94 & 8.77e-09 &  0.48 & 0.007 &&CL18 Z &1.10 & 1.34 & 1.06e-05 &  0.24 & 0.059 \\
CL9 X &0.92 & 1.00 & 1.05e-09 &  0.36 & 0.027 &&CL19 X &1.11 & 1.28 & 6.10e-07 &  0.25 & 0.033\\
CL9 Y &0.91 & 1.08 & 1.23e-07 &  0.37 & 0.008 &&CL19 Y &1.06 & 1.16 & 6.54e-07 &  0.33 & 0.026\\
CL9 Z &0.93 & 1.11 & 1.11e-07 &  0.39 & 0.027 && CL19 Z &1.10 & 1.29 & 1.56e-06 &  0.25 & 0.043\\
CL10 X &0.80 & 0.91 & 1.56e-07 &  0.39 & 0.008 && CL20 X &0.87 & 0.94 & 7.16e-08 &  0.42 & 0.011 \\
CL10 Y &0.84 & 1.04 & 4.72e-06 &  0.16 & 0.120 && CL20 Y &0.89 & 0.95 & 7.80e-08 &  0.42 & 0.011\\
CL10 Z &0.85 & 1.32 & 5.00e-06 &  0.22 & 0.104 && CL20 Z &0.88 & 0.92 & 1.44e-09 &  0.43 & 0.009\\

& & & & & & & & & & & & \\
\hline 
\end{tabular}
\label{tab:all}
\end{table*}

\acknowledgments
\bigskip

\begin{center}
{\bf Acknowledgment}
\end{center}
We are thankful to the referee who improved the manuscript through comments.
We are greatly indebted to Stefano Borgani, Dunja Fabjan, Annalisa Bonafede, Klaus Dolag, Giuseppe Murante, and Luca Tornatore who produced the simulations used in this work and to Emanuele Contini and Gabriella de Lucia to share the mass accretion history files. 
ER acknowledges useful discussions with Pasquale Mazzotta, Daisuke Nagai, Mary Hemmeter, Dai Phuc Po, and Dan Gantman on morphological estimators.
We gratefully acknowledge support from the National Science Foundation through grant AST-1210973, the Chandra Theory grant TM3-14008X and from contracts ASI-INAF I/023/05/0 and I/088/06/0.
ER and MM would like to thank the Michigan Center for Theoretical Physics for supporting the collaboration.

\bibliographystyle{../FILES/apj}
\bibliography{ref} 
\end{document}